\documentclass[aps,prd,showpacs,preprintnumbers]{revtex4-1}

\usepackage{amsmath,amsfonts,amssymb}
\usepackage{graphicx}
\usepackage{dcolumn}

\newcommand{\tr}{\mathop{\text{tr}}}

\begin{document}

\title{Cosmological models with running cosmological term and decaying dark matter}
\author{Marek Szyd{\l}owski}
\email{marek.szydlowski@uj.edu.pl}
\affiliation{Astronomical Observatory, Jagiellonian University, Orla 171, 30-244 Krakow, Poland}
\affiliation{Mark Kac Complex Systems Research Centre, Jagiellonian University, {\L}ojasiewicza 11, 30-348 Krak{\'o}w, Poland}
\author{Aleksander Stachowski}
\affiliation{Astronomical Observatory, Jagiellonian University, Orla 171, 30-244 Krakow, Poland}

\begin{abstract}
We investigate the dynamics of the generalized $\Lambda$CDM model, which the $\Lambda$ term is running with the cosmological time. The $\Lambda(t)$ term emerges from the covariant theory of the scalar field $\phi$ with the self-interacting potential $V(\phi)$. On the example of the model $\Lambda(t)=\Lambda_{\text{bare}}+\frac{\alpha^2}{t^2}$ we show the existence of a mechanism of the modification of the scaling law for energy density of dark matter: $\rho_{\text{dm}}\propto a^{-3+\lambda(t)}$. We discuss the evolution of $\Lambda(t)$ term and pointed out that during the cosmic evolution there is a long phase in which this term is approximately constant. This effect justifies Alcaniz and Lima's approach to $\Lambda(H)$ cosmologies. We also present the statistical analysis of both the $\Lambda(t)$CDM model with dark energy and decaying dark matter and the $\Lambda$CDM standard cosmological model. We divide the observational data into two groups: low $z$ data (SNIa, BAO, $H(z)$ and AP test) and high $z$ data (Planck, WP and lensing). While for the former we find the best fit value of the parameter $\lambda$ is positive ($\lambda=0.0338$, energy transfer is from the dark energy to dark matter sector), for the latter we find that $\lambda$ is $-0.0199$ which is an evidence that the energy transfer is from decaying dark matter. This disagreement correlates with estimated values of $H_0$ ($67.77$ km/(s Mpc) and $65.62$ km/(s Mpc) respectively). The decaying dark matter causes to lowering a mass of dark matter particles which are lighter than CDM particles and remain relativistic. The rate of the process of decaying matter is estimated. We show that in the models of decaying dark matter, the cosmological constant problem disappears naturally. The model with decaying dark matter possesses one parameter more but in light of the AIC it is better than the $\Lambda$CDM standard cosmological model.
\end{abstract}


\maketitle

\section{Introduction}
In cosmology, the standard cosmological model ($\Lambda$CDM model) is an effective theory which well describes the current Universe in the accelerating phase of the expansion. All the astronomical observations of supernovae SNIa and measurements of CMB favor this model over the alternatives but we are still looking for theoretical models to dethrone the $\Lambda$CDM model.

On the other hand the $\Lambda$CDM model has serious problems like the cosmological constant problem or the coincidence problem which are open and waiting for a solution. Among different propositions, it is an idea of introducing the running cosmological term \cite{Szydlowski:2015rga}. The most popular way of introducing a dynamical form of the cosmological term is a parametrization by the scalar field, i.e. $\Lambda\equiv \Lambda(\phi)$ or the Ricci scalar, i.e. $\Lambda\equiv \Lambda(R)$, where $R$ is the Ricci scalar. In the both mentioned cases, the covariance of field equation is not violated and $\Lambda\equiv \Lambda(t)$ relation emerges from covariant theories i.e. $\Lambda\equiv \Lambda(\phi(t))$. For example in paper \cite{Szydlowski:2015rga}, the relation $\Lambda=\Lambda_{\text{bare}}+\frac{\alpha^2}{t^2}$ is emerging from the theory of the scalar field with the self-interacting potential $V(\phi)$. It is interesting that this type of a $\Lambda(t)$ relation is supported by the non-critical string theory consideration \cite{Lopez:1995eb}. 

Two elements appear in the $\Lambda$CDM model, namely dark matter and dark energy. The main aim of observational cosmology is to constrain the density parameters for dark energy as well as dark matter. In the testing of the $\Lambda$CDM model, the idea of dark energy is usually separated from the dark matter problem, The latter is considered as the explanation of flat galactic curves. Of course the conception of dark matter is also needed for the consistency of the model of cosmological structures but the hypothesis of dark energy and dark matter should be tested not as a isolated hypothesis.

In this paper, we explore the $\Lambda(t)$CDM model with $\Lambda(t)=\Lambda_{\text{bare}}+\frac{\alpha^2}{t^2}$, where $t$ is the cosmological time for which we know an exact solution \cite{Szydlowski:2015rga}. This enables us to show the nontrivial interactions between the sectors of dark matter and dark energy. It would be demonstrated that the model, which is under consideration, constitutes the special case of models with the interaction \cite{Szydlowski:2015rga} term $Q=-\frac{d\Lambda(t)}{dt}$. We will be demonstrated that the time dependence of the $\Lambda$ term is responsible for the modification of the standard scaling law of dark matter $\rho_{\text{dm}}=\rho_{\text{dm},0} a^{-3}$, where $a$ is the scale factor \cite{Szydlowski:2015rga}. Wang and Meng \cite{Wang:2004cp} developed a phenomenological approach which is based on the modified matter scaling relation $\rho_{\text{m}}=\rho_{\text{m},0} a^{-3+\delta}$, where $\delta$ is the parameter which measures a deviation from the standard case of cold dark matter (CDM).

The both effect of the decaying $\Lambda$ term and the modification of the scaling relation are strictly related in our model. One can obtain that CDM particles dilute more slowly in comparison to the standard relation $\rho_{\text{m}} \propto a^{-3}$ in this model. In this context, the interesting approach was developed by Alcaniz and Lima \cite{Alcaniz:2005dg,Graef:2013iia}. The coupling parameter $\delta$ is also a subject of testing using astronomical data \cite{Bessada:2013maa,Szydlowski:2015bwa,Goncalves:2015eaa}.

In this paper, due to it is known the exact solutions of our model it is possible to check how it works the model and one can strictly constrain the model parameters \cite{Szydlowski:2015rga}. 

We estimate the value of $\lambda(t) \colon \rho_{\text{dm}} = \rho_{\text{dm},0} a^{-3+\lambda(t)}$ where $\rho_{\text{dm}}$ is energy density of dark matter. In our methodology we consider the astronomical data as two separate data sets with respect to redshift $z$. The first group of low redshift consists of SNIa, BAO, $H(z)$ and the AP test, and the second group is composed of Planck, WP and lensing data. The case of decaying dark matter effect is only relevant for high redshift data. This process takes place in the early Universe.

We also analyze the model under considerations in details. In this analysis the model with $\Lambda(t)=\Lambda_{\text{bare}}+\frac{\alpha^2}{t^2}$ is our case study. For this model we show the terms $\lambda(t)$, $\delta(t)$ are slow-changing with respect to the cosmological time and it justified to treat them as constants.

The organization of the text is following. In Section~\ref{sec:2}, we present the model with $\Lambda(t)=\Lambda_{\text{bare}}+\frac{\alpha^2}{t^2}$ and its interpretation in scalar field cosmology with $\phi(t)$ and with the potential $V(\phi)$. In Section~\ref{sec:3}, it is demonstrated how $\Lambda(t)$CDM cosmologies can be interpolated as interacting cosmologies with the interacting term $Q=-\frac{d\Lambda(t)}{dt}$ and how they solve the coincidence problem \cite{delCampo:2015vha}. In Section~\ref{sec:4}, we present some results of the statistical estimations of the model parameters obtained from some astronomical data. Finally the conclusion are presented in Section~\ref{sec:5}.

\section{$\Lambda(t)$CDM cosmology with $\Lambda=\Lambda_{\text{bare}}+\frac{\alpha^2}{t^2}$} \label{sec:2}

Let us consider about the flat cosmological model with homogeneity and isotropy (the Robertson-Walker symmetry). The source of gravity is in the time dependent cosmological term and matter is in the form of a perfect fluid with energy density $\rho_{\text{m}}=\rho_{\text{m}} (t)$, where $t$ is the cosmological time. The cosmic evolution is determined by the Einstein equations which admit the Friedmann first integral in the form
\begin{equation}
    3H^2(t)=\rho_{\text{m}} (t)+\Lambda_{\text{bare}}+\frac{\alpha^2}{t^2},\label{friedmann}
\end{equation}
where $H(t)=\frac{d\log a}{dt}$ is the Hubble function and $a(t)$ is the scale factor and $\alpha^2 \in \mathbb{R}$ is a real dimensionless parameter. The sign of $\alpha^2$ depends of the type of particle and the distribution of their energy. In the generic case the Breit-Wigner distribution gives rise the negative sign of $\alpha^2$ \cite{Urbanowski:2013tfa}. Note that this parametrization is distinguished by a dimensional analysis because a dimension of $H^2$ should coincide with a dimension of a time dependent part of $\Lambda(t)$. 

It is assumed that the energy-momentum tensor for all fluids in the form of perfect fluid satisfies the conservation condition
\begin{equation}
    T^{\alpha \beta}_{;\alpha}=0,
\end{equation}
where $T^{\alpha \beta}=T_m^{\alpha \beta}+\Lambda(t)g^{\alpha\beta}$. The consequence of this relation is that
\begin{equation}
    \dot{\rho}_{\text{m}}+3H\rho_{\text{m}} = -\frac{d\Lambda}{dt}. \label{con}
\end{equation}
The cosmic evolution is governed by the second order acceleration equation
\begin{equation}
    \dot H=-H^2-\frac{1}{6}(\rho_{\text{eff}}+3p_{\text{eff}}),
\end{equation}
where $\rho_{\text{eff}}$ and $p_{\text{eff}}$ are effective energy density of all fluids and pressure respectively.
In the model under the consideration we have
\begin{equation}
    \rho_{\text{eff}}=\rho_{\text{m}}+\rho_{\Lambda},
\end{equation}
\begin{equation}
    p_{\text{eff}}=p_{\text{m}}-\rho_{\Lambda},
\end{equation}
where $p_{\text{m}}=0$, $\rho_{\Lambda}=\Lambda_{\text{bare}}+\frac{\alpha^2}{t^2}$ and $\alpha^2$ is a real number.

For this case the exact solution of (\ref{friedmann}) and (\ref{con}) for the Hubble parameter $h\equiv \frac{H}{H_0}$ can be obtained in terms of Bessel functions
\begin{equation}
   h(t)=\frac{1-2n}{3H_0 t}+\sqrt {-\Omega_{\Lambda,0}}\frac{I_{n-1}\left(\frac{3\sqrt {\Omega_{\Lambda,0}} H_0}{2}t \right)}{
I_n \left(\frac{3\sqrt {\Omega_{\Lambda,0}} H_0}{2}t \right)}.
 \label{hubble}
\end{equation}
where $H_0$ is the present value of the Hubble constant, $\Omega_{\Lambda,0}=\frac{\Lambda_{\text{bare}}}{3H_0^2}$, $\Omega_{\alpha,0}=\frac{\alpha^2 T_0^2}{3 H_0^2}$, $T_0$ is the present age of the Universe $T_0=\int^{T_0}_0 dt$ and $n=\frac{1}{2}\sqrt{1+9\Omega_{\alpha,0}T_0^2 H_0^2}$ is the index of the Bessel function. From (\ref{hubble}), the expression for the scale factor can be obtained in the simple form
\begin{equation}
	a(t)=C_2 \left[\sqrt t \left(
I_n \left(\frac{3\sqrt {\Omega_{\Lambda,0} }H_0}{2}t\right)\right)\right]^\frac{2}{3}.
\end{equation}
The inverse formula for $t(a)$ is given by
\begin{equation}
    t(a)=\frac{2}{3i\sqrt{\Omega_{\Lambda,0} H_0}}S_{n-\frac{1}{2}}^{-1}\left(\frac{\sqrt{2\pi} i^{n-1}}{3\sqrt{\Omega_{\Lambda,0}} H_0 C_2}a^\frac{3}{2}\right),
\end{equation}
where $S_n (x)$ is the Riccati-Bessel function $S_n(x)=\sqrt{\frac{\pi x}{2}}J_{n+\frac{1}{2}}(x)$.

Finally the exact formula for total mass $\rho_m(t)= \rho_{\text{dm}}(t) + \rho_{\text{b}}(t)$ is given in the form
\begin{equation}
\rho_m=-3H_0^2\left(\Omega_{\Lambda,0}+\frac{\Omega_{\alpha,0}T^2_0}{t^2}-\left(\frac{1-2n}{3H_0 t}+\sqrt {\Omega_{\Lambda,0}}\frac{I_{n-1}\left(\frac{3\sqrt {\Omega_{\Lambda,0}} H_0}{2}t \right)}{
I_n \left(\frac{3\sqrt {\Omega_{\Lambda,0}} H_0}{2}t \right)}\right)^2\right).
\end{equation}

\begin{figure}[ht]
\centering
\includegraphics[scale=0.4]{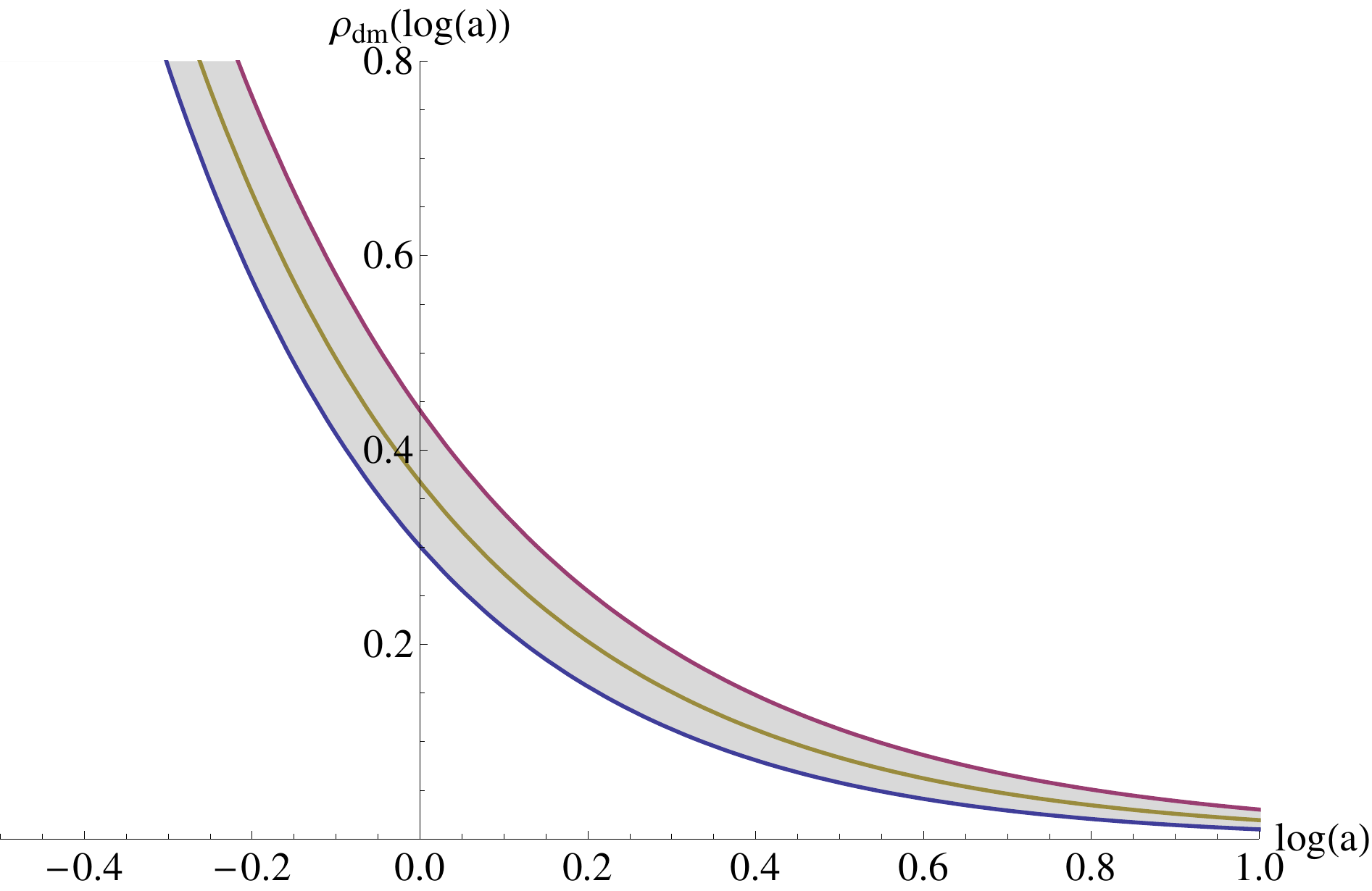}
\caption{A diagram of the evolution of $\rho_{\text{dm}}(\log(a))$. The top thick line represents the evolution of $\rho_{\text{dm}}(\log(a))$ for $H_0=67.64$ km/(s Mpc), $\Omega_{\text{m},0}=0.3696$ and $\Omega_{\alpha^2}=0.0392$. The bottom thick line represents the evolution of $\rho_{\text{dm}}(\log(a))$ for $H_0=68.13$ km/(s Mpc), $\Omega_{\text{m},0}=0.2646$ and $\Omega_{\alpha^2}=-0.0258$. The medium line represents the best fit (see Table~\ref{table:1}). The gray region is the $2\sigma$ uncertainties. We assumed $8\pi G=1$ and we choose $\text{km}^2/(100^2 \text{Mpc}^2 s^2)$ as a unit of $\rho_{\text{dm}}(\log(a))$.} 
\label{fig:4}
\end{figure}

\begin{figure}[ht]
\centering
\includegraphics[scale=0.4]{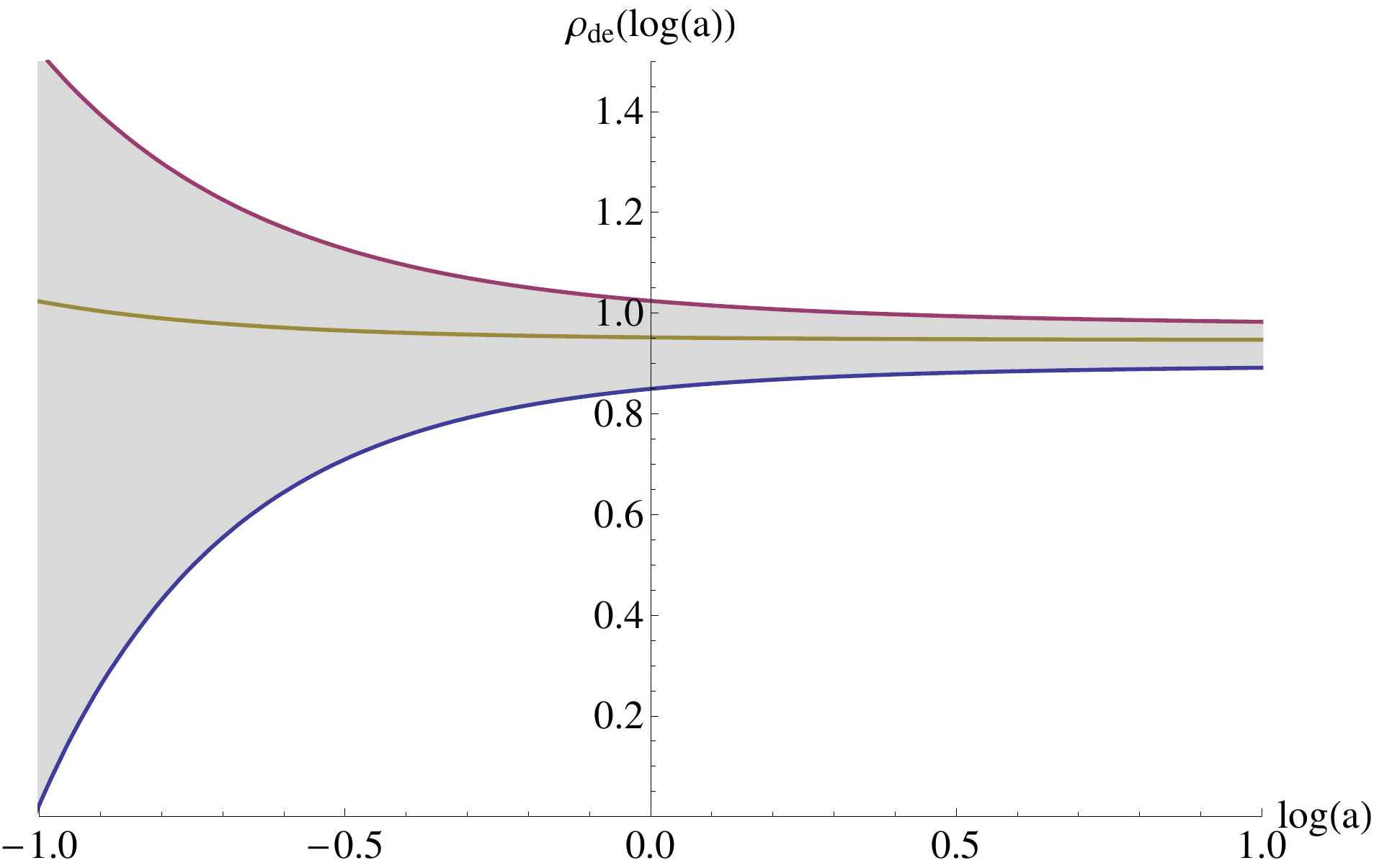}
\caption{A diagram of the evolution of $\rho_{\text{de}}(\log(a))$. The bottom thick line represents the evolution of $\rho_{\text{de}}(\log(a))$ for $H_0=67.64$ km/(s Mpc), $\Omega_{\text{m},0}=0.3696$ and $\Omega_{\alpha^2}=0.0392$. The top thick line represents the evolution of $\rho_{\text{de}}(\log(a))$ for $H_0=68.13$ km/(s Mpc), $\Omega_{\text{m},0}=0.2646$ and $\Omega_{\alpha^2}=-0.0258$. The medium line represents the best fit (see Table~\ref{table:1}). The gray region is the $2\sigma$ uncertainties. We assumed $8\pi G=1$ and we choose $\text{km}^2/(100^2 \text{Mpc}^2 s^2)$ as a unit of $\rho_{\text{de}}(\log(a))$.} 
\label{fig:5}
\end{figure}

\begin{figure}[ht]
\centering
\includegraphics[scale=0.4]{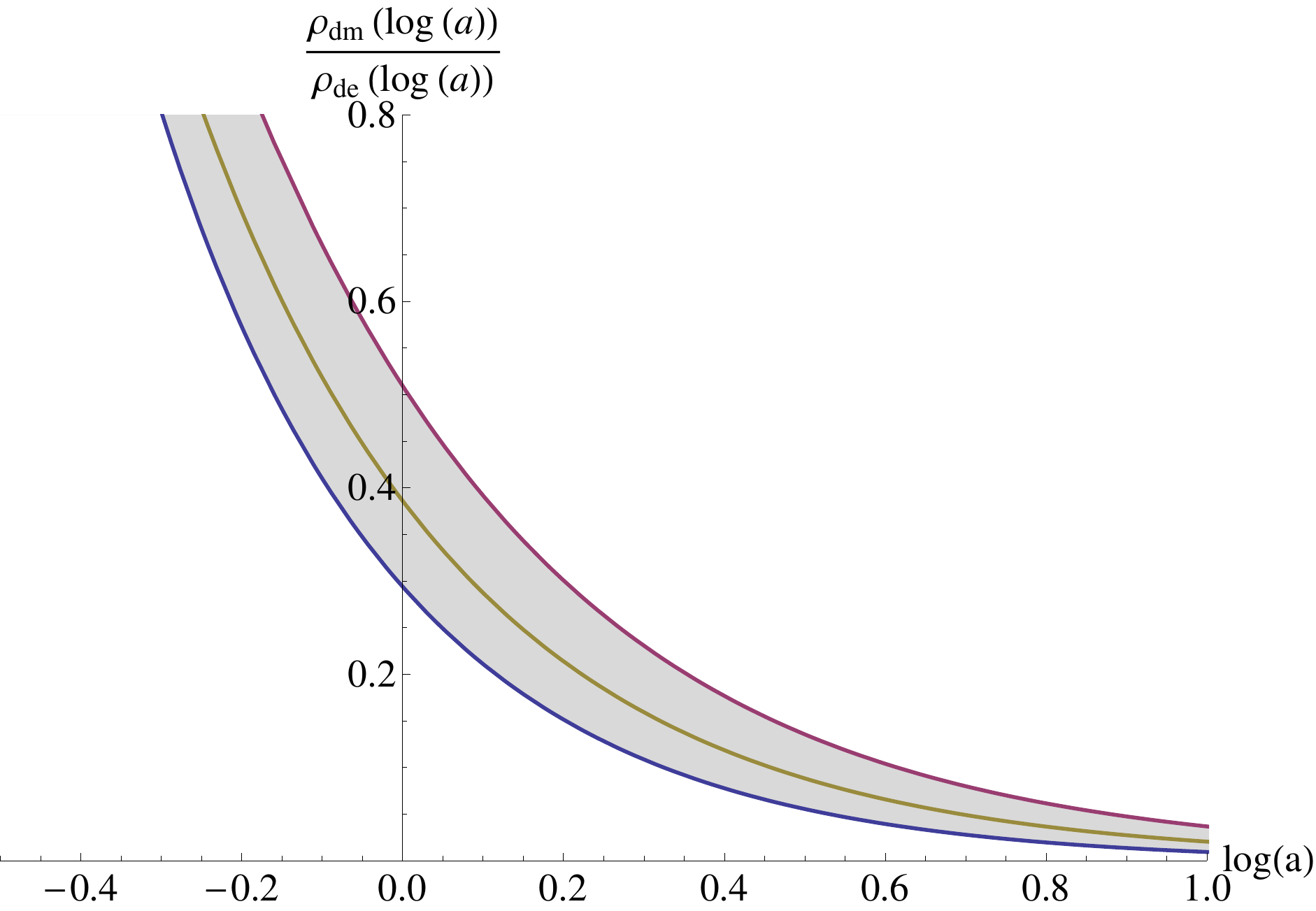}
\caption{A diagram of the evolution of $\frac{\rho_{\text{dm}}(\log(a))}{\rho_{\text{de}}(\log(a))}$. The top thick line represents the evolution of $\frac{\rho_{\text{dm}}(\log(a))}{\rho_{\text{de}}(\log(a))}$ for $H_0=67.64$ km/(s Mpc), $\Omega_{\text{m},0}=0.3696$ and $\Omega_{\alpha^2}=0.0392$. The bottom thick line represents the evolution of $\frac{\rho_{\text{dm}}(\log(a))}{\rho_{\text{de}}(\log(a))}$ for $H_0=68.13$ km/(s Mpc), $\Omega_{\text{m},0}=0.2646$ and $\Omega_{\alpha^2}=-0.0258$. The medium line represents the best fit (see Table~\ref{table:1}). The gray region is the $2\sigma$ uncertainties.} 
\label{fig:6}
\end{figure}

The diagram of $\rho_{\text{dm}}$, $\rho_{\text{de}}$ and $\frac{\rho_{\text{dm}}(\log(a))}{\rho_{\text{de}}(\log(a))}$ as a function of $\log a$ obtained for low $z$ data is presented in Fig.~\ref{fig:4}, \ref{fig:5} and~\ref{fig:6}. Note that at the present epoch ($\log(a)=0$) both energy densities of dark matter and dark energy are of the same order (Fig.~\ref{fig:5}).

While the relation $\Lambda = \Lambda(t)$ violates the covariance of the general relativity Lagrangian, it can be simply demonstrated that such a relation can emerge from the covariant theory of scalar fields. This relation reveals as a time-dependent first integral of equation of motion. 

The action of general relativity for a perfect fluid has the following form
\begin{equation}
S=\int\sqrt{-g}(R+\mathcal{L}_m)d^4 x,
\end{equation}
where $R$ is the Ricci scalar, $\mathcal{L}_m=-\rho\left(1+\int \frac{p(\rho)}{\rho^2}d\rho\right)$ \cite{Minazzoli:2012md} and $g_{\mu\nu}$ is the metric tensor. The signature of $g_{\mu\nu}$ is chosen as $(+,-,-,-)$.

For the Friedmann-Lemaitre-Robertson-Walker (FLRW) metric without the curvature the Ricci scalar is expressed by $R=6\left(\frac{\ddot{a}}{a}+\frac{\dot{a}^2}{a^2}\right)$. The Einstein equations are consequence of the variation of the Langragian $\mathcal{L}$. The Einstein equations for dust and a minimal coupling scalar field are the following
\begin{equation}
    3H^2=\rho \label{eqf}
\end{equation}
and
\begin{equation}
    \frac{\ddot a}{a}=-\frac{1}{6}(\rho+3p). \label{eqd}
\end{equation}
We assume that $\rho=\rho_{\text{m}}+\rho_{\text{de}}$ and an interaction $Q=\frac{2\alpha^2}{t^3}$ between $\rho_{\text{m}}$ and $\rho_{\text{de}}$ and we assume the state equation for $\rho_{\text{m}}$ as $p_{\text{m}}=0$ and for $\rho_{\text{de}}$ as $p_{\text{de}}=-\rho_{\text{de}}$. Now we choice a new parametrization of $\rho_{\text{de}}$ as $\rho_{\text{de}}=\frac{\dot\phi^2}{2}+V(\phi)$, where $\phi=\phi(t)$. Then (\ref{eqf}) and (\ref{eqd}) have the following form
\begin{equation}
    3H^2=\rho_m+\frac{\dot{\phi}^2}{2}+V(\phi)
\end{equation}
and
\begin{equation}
     \frac{\ddot a}{a}=-\frac{1}{6}\left(\rho-\dot\phi^2-2V(\phi)\right).
\end{equation}

The conservation equation is in the form
\begin{equation}
    \dot\rho+3H(\rho+p)=\dot\rho_{\text{m}}+\dot\rho_{\text{de}}+3H\rho_{\text{m}}=0.
\end{equation}
Because there is an interaction $Q=\frac{2\alpha^2}{t^3}$ between $\rho_{\text{m}}$ and $\rho_{\text{de}}$ then the conservation equation gives two equations
\begin{equation}
     \dot\rho_{\text{m}}+3H\rho_{\text{m}}=\frac{2\alpha^2}{t^3}
\end{equation}
and
\begin{equation}
\dot\rho_{\text{de}}=\dot\phi\ddot\phi+\frac{dV(\phi)}{d\phi}\dot\phi=-\frac{2\alpha^2}{t^3}. \label{conphi}
\end{equation}
The first integral of (\ref{conphi}) is $\frac{\dot\phi^2}{2}+V(\phi)=\frac{\alpha^2}{t^2}+\Lambda_{\text{bare}}$. The potential $V(\phi)$ can be of any form. So if we want to find an exact solution for $\phi(t)$ we should first assume the form of the potential $V(\phi)$.

\section{How $\Lambda(t)$CDM model modifies the scaling relation for dark matter} \label{sec:3}

The existence of dark matter in the Universe is motivated by modern astrophysical observations as well as cosmological observations. From observations of rotation curves of spiral galaxies, masses of infracluster gas, gravitational lensing of clusters of galaxies to cosmological observations of the cosmic microwave background anisotropy and large scale structures we obtain strong evidences of dark matter.

Because models of nucleosynthesis of the early Universe are strongly restricted the fraction of baryons, we conclude that the nature of dark matter cannot be baryonic matter. On the other hand we imagine that particles of dark matter form a part of standard model (SM) of particles physics. There are many candidates for particles of dark matter e.g. WIMPs. Lately sterile neutrinos have been also postulated in this context \cite{Boyarsky:2009ix,Motohashi:2012wc}. The interesting approach is a search of photon emission from the decay or the annihilation of dark matter particles through the astrophysical observations of X-ray regions \cite{Abazajian:2001vt,Boyarsky:2014jta,Dolgov:2000ew}. For example the radiatively decaying dark matter particles as sterile neutrinos have been searched using X-ray observations \cite{Sekiya:2015jsa}.

Let us consider the $\Lambda$CDM model which describes a homogeneous and isotropic universe which consists of baryonic and dark matter and dark energy. Let us assume an interaction in the dark sectors. Then the conservation equations have the following form
\begin{align}
    \dot\rho_{\text{b}}+3H\rho_{\text{b}} &= 0,\label{con1} \\
    \dot\rho_{\text{dm}}+3H\rho_{\text{dm}} &= Q,\label{con2} \\
    \dot\rho_{\text{de}}+3H\rho_{\text{de}} &= -Q,
\end{align}
where $\rho_{\text{b}}$ is baryonic matter density, $\rho_{\text{dm}}$ is dark matter density and $\rho_{\text{de}}$ is dark energy density \cite{delCampo:2015vha}. $Q$ describes the interaction in the dark sector.

Let $\rho_{\text{m}}=\rho_{\text{b}}+\rho_{\text{dm}}$ then (\ref{con1}) and (\ref{con2}) give
\begin{equation}
    \dot\rho_{\text{m}}+3H\rho_{\text{m}}=Q \label{con4}.
\end{equation}
For model with $\Lambda(t)=\Lambda_{\text{bare}}+\frac{\alpha^2}{t^2}$ the conservation equation has the form $\dot\rho_{\text{m}}+3H\rho_{\text{m}} = -\frac{d\Lambda(t)}{dt}$. So this model can be interpreted as the special case of model with the interacting in the dark sectors. In this model $Q = -\frac{d\Lambda(t)}{dt}=\frac{2\alpha^2}{t^3}$.

\begin{figure}[ht]
\centering
\includegraphics[scale=0.4]{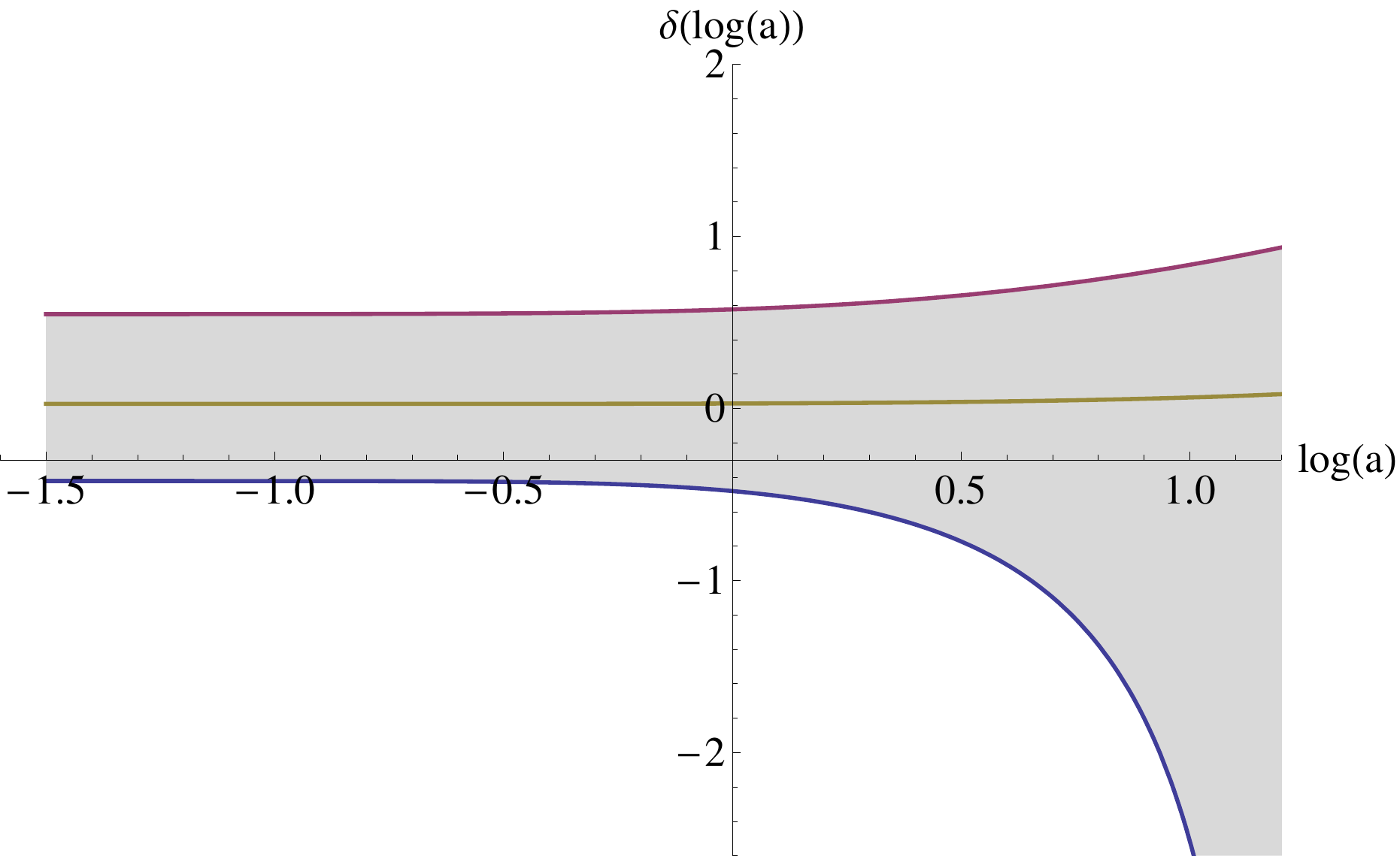}
\caption{A diagram of the evolution of $\delta(\log(a))$. The top thick line represents the evolution of $\delta(\log(a))$ for $H_0=68.67$ km/(s Mpc), $\Omega_{\text{m},0}=0.3311$ and $\Omega_{\alpha^2}=0.1034$. The bottom thick line represents the evolution of $\delta(\log(a))$ for $H_0=67.34$ km/(s Mpc), $\Omega_{\text{m},0}=0.2975$ and $\Omega_{\alpha^2}=-0.0626$.  The medium line represents the best fit (see Table~\ref{table:1}). The gray region is the $2\sigma$ uncertainties. } 
\label{fig:1}
\end{figure}

\begin{figure}[ht]
\centering
\includegraphics[scale=0.4]{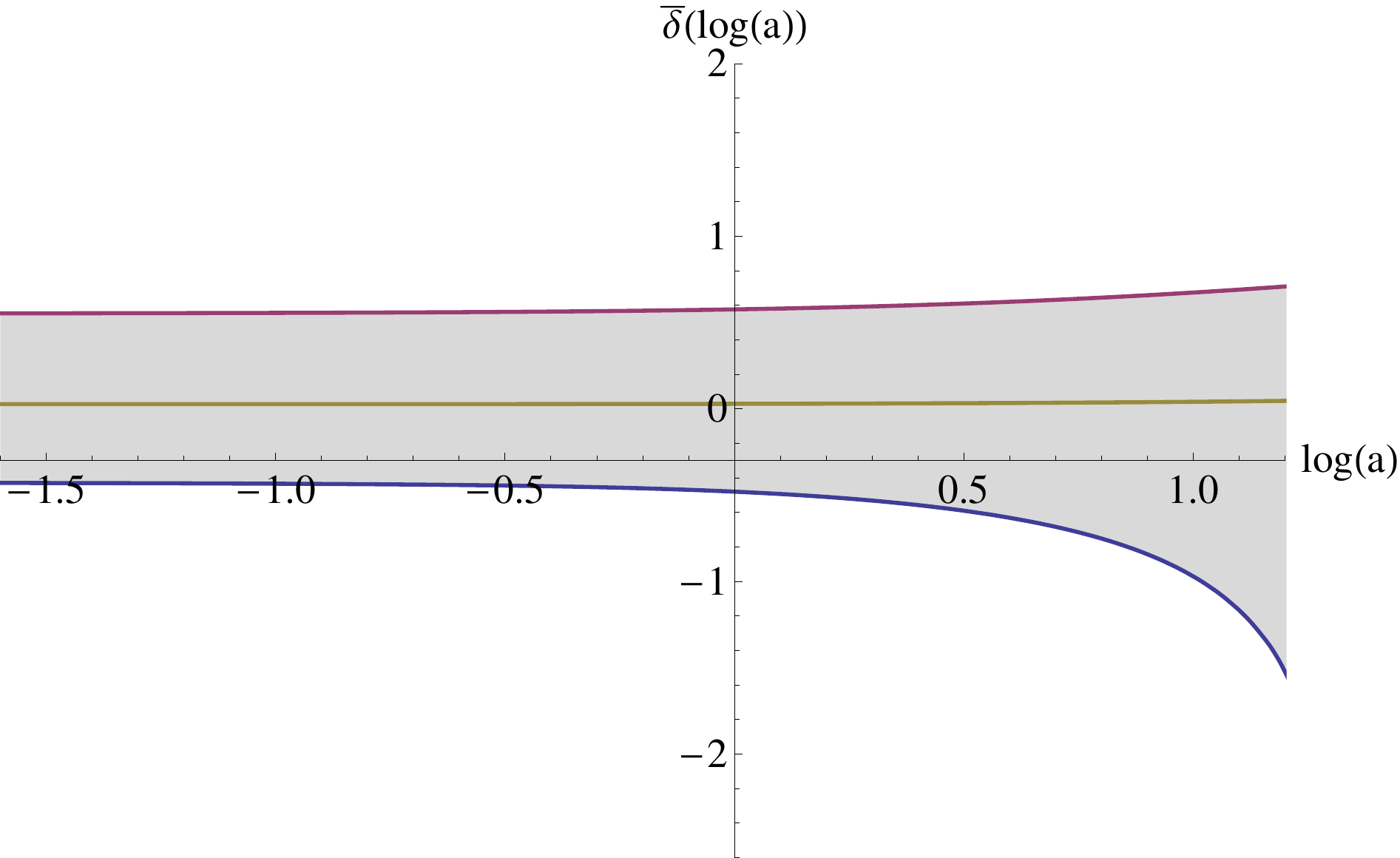}
\caption{A diagram of the evolution of $\bar\delta(\log(a))$. The top thick line represents the evolution of $\bar\delta(\log(a))$ for $H_0=68.67$ km/(s Mpc), $\Omega_{\text{m},0}=0.3311$ and $\Omega_\alpha^2=0.1034$. The bottom thick line represents the evolution of $\bar\delta(\log(a))$ for $H_0=67.34$ km/(s Mpc), $\Omega_{\text{m},0}=0.2975$ and $\Omega_\alpha^2=-0.0626$. The medium line represents the best fit (see Table~\ref{table:1}). The gray region is the $2\sigma$ uncertainties. Note that if $\rho_{\text{dm}}=0$ for $\alpha^2<0$, i.e. whole dark matter decays then we have the $\Lambda$CDM model with baryonic matter.} 
\label{fig:2}
\end{figure}

Equation (\ref{con4}) for $Q=\frac{2\alpha^2}{t^3}$ can be rewritten as 
$ \dot\rho_{\text{m}}+3H\rho_{\text{m}} = H\rho_{\text{m}}\frac{2\alpha^2}{t^3 H\rho_{\text{m}}}$ or
\begin{equation}
    \frac{d\rho_{\text{m}}}{\rho_{\text{m}}}=\frac{da}{a}\left(-3+\frac{2\alpha^2}{t^3 H\rho_{\text{m}}}\right). \label{con5}
\end{equation}
The solution of equation (\ref{con5}) is
\begin{equation}
    \rho_{\text{m}}=\rho_{\text{m},0} a^{-3+\bar\delta(t)}, \label{solrho}
\end{equation}
where $\bar\delta=\frac{1}{\log a}\int\delta(t)d \log a$, where $\delta(t)=\frac{2\alpha^2}{t^3 H(t)\rho_{\text{m}}(t)}$. $Q$ can be written as $Q=\delta(t)H\rho_{\text{m}}$. The evolution of $\delta(\log(a))$ and $\bar\delta(\log(a))$ is presented in Fig.~\ref{fig:1}, and \ref{fig:2}. One can observe that $\bar\delta(t)$ and $\delta(t)$ is constant since the initial singularity to the present epoch.

If $\delta(t)$ is a slowly changing function than $\bar\delta(t)=\delta(t)=\delta$ and (\ref{solrho}) has the following form
\begin{equation}
     \rho_{\text{m}}=\rho_{\text{m},0} a^{-3+\delta}.
\end{equation}
In this case $Q=\delta H\rho_{\text{m}}$.

The early time approximation for $\delta(t)$ is
\begin{equation}
    \delta(t)=\frac{18\alpha^2}{4\sqrt{1+3\alpha^2}+6\alpha^2+4}.
\end{equation}
If $\delta(t)=\delta=\text{const}$ we can easily find that
\begin{equation}
    a=a_0 t^{\frac{2}{3-\delta}}.
\end{equation}
Let us note that during the regime $\delta=\text{const}$ both the time dependent parametrization and quadratic parametrization $\Lambda(H) = \Lambda_{\text{bare}} + \beta H^2$ coincide in Alcaniz and Lima approach \cite{Alcaniz:2005dg}.

\begin{figure}[ht]
\centering
\includegraphics[scale=0.4]{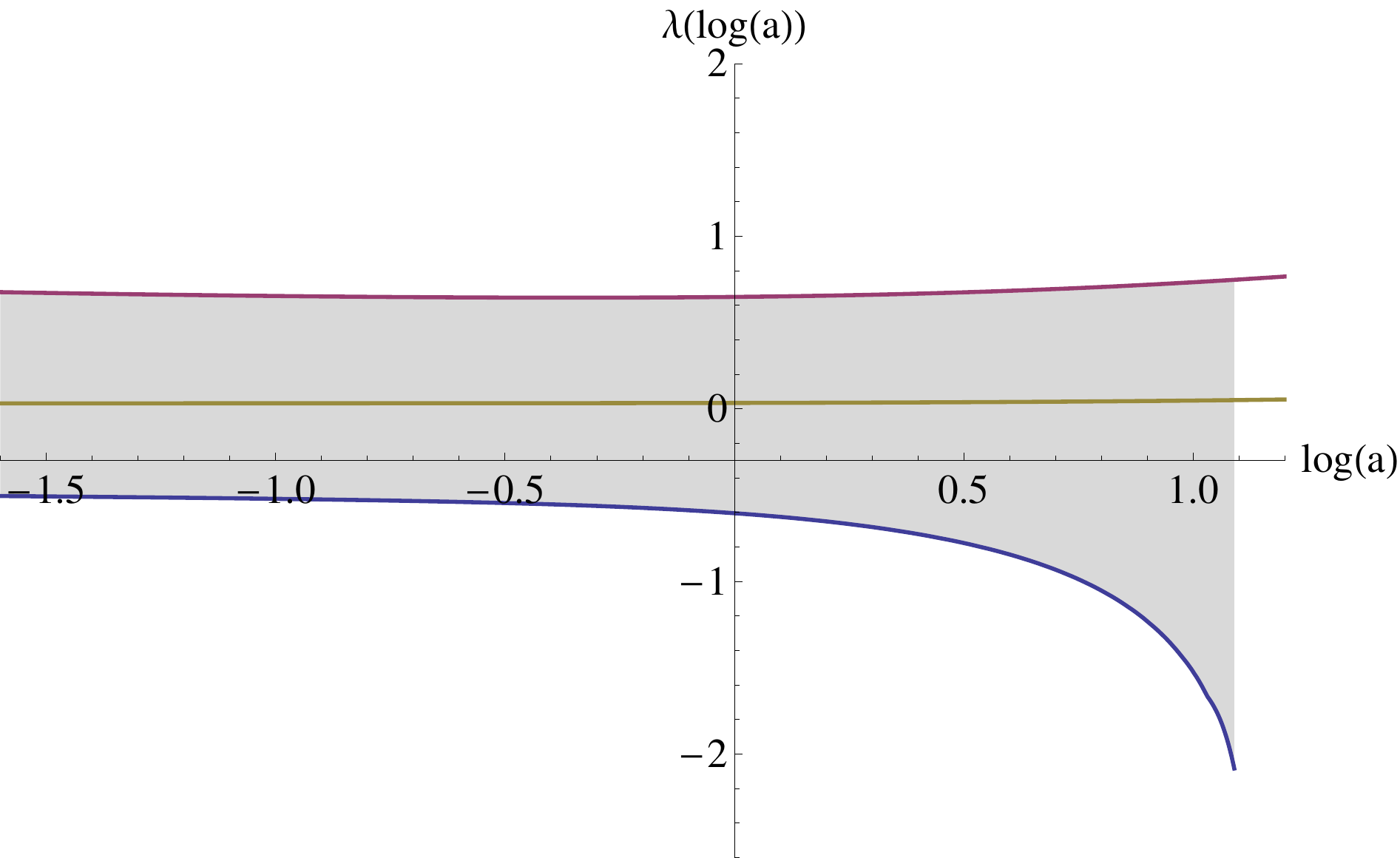}
\caption{A diagram of the evolution of $\lambda(\log(a))$.  The top thick line represents the evolution of $\lambda(\log(a))$ for $H_0=68.67$ km/(s Mpc), $\Omega_{\text{m},0}=0.3311$ and $\Omega_\alpha^2=0.1034$. The bottom thick line represents the evolution of $\lambda(\log(a))$ for $H_0=67.34$ km/(s Mpc), $\Omega_{\text{m},0}=0.2975$ and $\Omega_\alpha^2=-0.0626$. The medium line represents the best fit (see Table~\ref{table:1}). The gray region is the $2\sigma$ uncertainties.} 
\label{fig:3}
\end{figure}

We can rewrite $\rho_{\text{dm}}$ as
\begin{equation}
    \rho_{\text{dm}}=\rho_{\text{dm},0}a^{-3+\lambda(a)}, \label{rhodm}
\end{equation}
where $\lambda(t)=\frac{1}{\log a} \log\frac{\Omega_{\text{m},0}a^{\bar\delta(t)}-\Omega_{\text{b}}}{\Omega_{\text{m},0}-\Omega_{\text{b}}}$. For the present epoch we can approximate $\lambda(t)$ as $\lambda(t)=\lambda=\text{const}$. So in the present epoch $\rho_{\text{dm}}=\rho_{\text{dm},0}a^{-3+\lambda}$. In the consequence, the Friedmann equation can be written as $3H^2=\rho_{\text{b}} a^{-3}+\rho_{\text{dm}}a^{-3+\lambda}+\Lambda_{\text{bare}}+\frac{\alpha^2}{t^2}$. The evolution of $\lambda(\log(a))$ is presented in Fig.~\ref{fig:3}. One can observe that $\lambda(t)$ is constant since the initial singularity to the present epoch.

Following Amendola and others \cite{Amendola:2006dg,Majerotto:2004ji,Rosenfeld:2005pw} the mass of dark particles can be parametrized by the scale factor as
\begin{equation}
    m(a)=m_{0}\exp\int^a \kappa(a')d(\log a'), \label{m}
\end{equation}
where $m_0$ is representing of mass of dark matter, $\kappa=\frac{d\log m}{d\log a}$. We consider the mass $m(a)$ as an effective mass of particles in a comoving volume. 

In Amendola et al. \cite{Amendola:2006dg} the parameter $\kappa(a)$ is assumed as a constant. This simplifying assumption has physical justification as it will be demonstrated in the further dynamical analysis of the model with decaying dark matter. Equation (\ref{m}) can be simply obtained from (\ref{rhodm}) because $e=a^{\frac{1}{\log a}}$ and $m(a)=a^3 \rho(a)$. Then $\lambda(a)= \frac{1}{\log(a)} \int^a \kappa(a')d(\log a')$. For illustration the rate of dark matter decaying process it would be useful to define the parameter $\beta$ 
\begin{equation}
\beta = 2^{\frac{\delta -3}{2\lambda}}.
\end{equation}
If $\lambda(a)=\text{const}$ then equation (\ref{m}) has the equivalent form $m(t)=m_0 a_0^\lambda \exp^{-\log \frac{2t}{\beta}}$, where $\beta=2^{\frac{\delta-3}{2\lambda}}$.

Let consider the number of dark matter particles $N(t)$ where $t$ is the cosmological time. Than a half of the number of these particle $N(t)/2$ is reached at the moment of time $\beta t$.

For small value of $\delta \ll 3$ the parameter $\beta$ can be approximated by $\beta = 2^{\frac{-3}{2\lambda}}$. For example if we put $\lambda = 0.06$ than the number of dark matter particles decreases to half after about $33.5 \cdot 10^{6} t_0$ starting at the time $t_0$.

\begin{figure}[ht]
\centering
\includegraphics[scale=0.4]{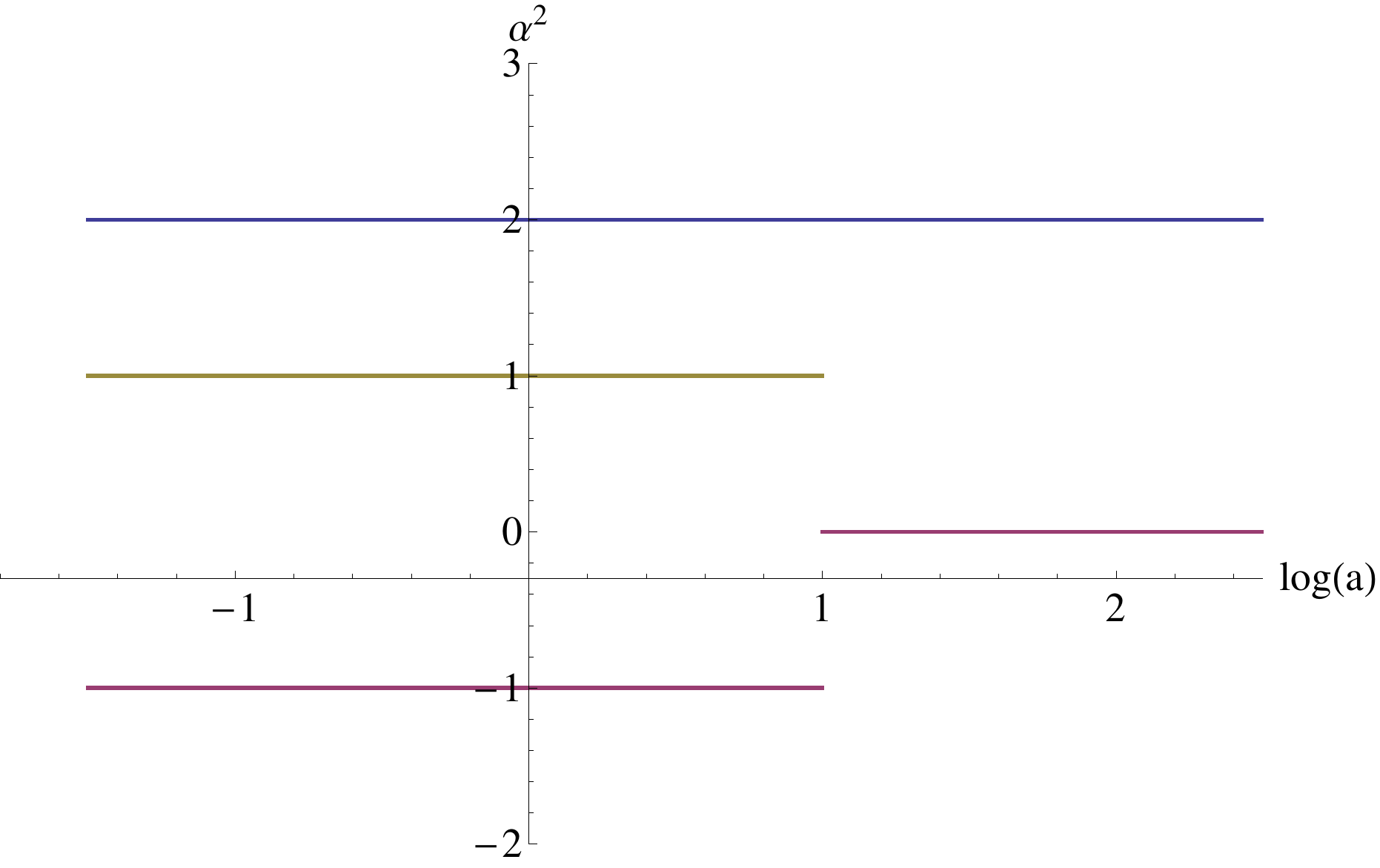}
\caption{A diagram of the switching phenomenon of $\alpha^2$. The bottom left line represents $\alpha^2$ for the first fraction of particles and the top left  line represents $\alpha^2$ for the second fraction of particles before switching. The medium left line represents effective $\alpha^2$. After switching the bottom line represents $\alpha^2$ of the first fraction which is equal zero.} 
\label{fig:7}
\end{figure}

In Fig.~\ref{fig:6} one can see that a quotient $\rho_{\text{dm}}/\rho_{\text{de}}$ decreases with the scale factor and remains of the same order for today ($\log a = 0$). Therefore the coincidence problem is solved.

If the dark matter sector consists of two fractions of particles than $\alpha^2=\alpha_{\text{eff}}^2=\alpha_1^2+\alpha_2^2$, where $\alpha_1^2$ is a value of $\alpha^2$ for the first fraction of particles and $\alpha_1^2$ is value of $\alpha^2$ for the second fraction of particles. If particles of the first fraction is decayed than $\alpha^2$ for this fraction should be equal zero and the value of $\alpha_{\text{eff}}^2$ is switched. Than $\alpha_{\text{eff}}^2=\alpha_2^2$. This phenomenon is demonstrated in Fig.~\ref{fig:7}.

\section{Statistical analysis of the model} \label{sec:4}

In this section, we present a statistical analysis of the model parameters such as $H_0$, $\Omega_{\text{dm},0}$ and $\lambda$. We are using the SNIa, BAO, CMB observations, measurements of $H(z)$ for galaxies and the Alcock-Paczy{\'n}ski test.

We use the data from Union 2.1 which is the sample of 580 supernovae \cite{Suzuki:2011hu}. The likelihood function for SNIa is
\begin{equation}
    \log L_{\text{SNIa}} = -\frac{1}{2} [A - B^2/C + \log(C/(2 \pi))],
\end{equation}
where $A= (\mathbf{\mu}^{\text{obs}}-\mathbf{\mu}^{\text{th}})\mathbb{C}^{-1}(\mathbf{\mu}^{\text{obs}}-\mathbf{\mu}^{\text{th}})$, 
$B= \mathbb{C}^{-1}(\mathbf{\mu}^{\text{obs}}-\mathbf{\mu}^{\text{th}})$, $C=\tr \mathbb{C}^{-1}$ and $\mathbb{C}$ is a covariance matrix for SNIa.
The distance modulus is $\mu^{\text{obs}}=m-M$ (where $m$ is the apparent magnitude and  $M$ is the absolute magnitude of SNIa) and $\mu^{\text{th}} = 5 \log_{10} D_L +25$ (where the luminosity distance is $D_L= c(1+z) \int_{0}^{z} \frac{d z'}{H(z)}$).

We use Sloan Digital Sky Survey Release 7 (SDSS DR7) dataset at $z = 0.275$ \cite{Percival:2009xn}, 6dF Galaxy Redshift Survey measurements at redshift $z = 0.1$ \cite{Beutler:2011hx}, the BOSS DR 9 measurements at $z = 0.57$ \cite{Anderson:2012sa}, and WiggleZ measurements at redshift $z = 0.44, 0.60, 0.73$ \cite{Blake:2012pj}.
The likelihood function is given by
\begin{equation}
\log L_{\text{BAO}} = - \frac{1}{2}\left(\mathbf{d}^{\text{obs}}-\frac{r_s(z_d)}{D_V(\mathbf{z})}\right)\mathbb{C}^{-1}\left(d_i^{\text{obs}}-\frac{r_s(z_d)}{D_V(\mathbf{z})}\right), 
\end{equation}
where $r_s(z_d)$ is the sound horizon at the drag epoch \cite{Eisenstein:1997ik}.

The likelihood function for the Planck observations of cosmic microwave background (CMB) radiation \cite{Ade:2013zuv}, lensing from the Planck and low-$\ell$ polarization from the WMAP (WP) has the form
\begin{equation}
\log L_{\text{CMB}+\text{lensing}+\text{WP}} = - \frac{1}{2}  (\mathbf{x}^{\text{th}}-\mathbf{x}^{\text{obs}}) \mathbb{C}^{-1} (\mathbf{x}^{\text{th}}-\mathbf{x}^{\text{obs}}),
\end{equation}
where $\mathbb{C}$ is the covariance matrix with the errors, $\mathbf{x}$ is a vector of the acoustic scale $l_{A}$, the shift parameter $R$ and $\Omega_{b}h^2$ where
\begin{align}
l_A &= \frac{\pi}{r_s(z^{*})} c \int_{0}^{z^{*}} \frac{dz'}{H(z')} \\
R &= \sqrt{\Omega_{\text{m,0}} H_0^2} \int_{0}^{z^{*}} \frac{dz'}{H(z')},
\end{align}
where $z^{*}$ is the redshift of the photon-decoupling surface.

The likelihood function for the Alcock-Paczynski test \cite{Alcock:1979mp,Lopez-Corredoira:2013lca} has the following form
\begin{equation}
\log L_{AP} =  - \frac{1}{2} \sum_i \frac{\left( AP^{th}(z_i)-AP^{obs}(z_i) \right)^2}{\sigma^2}.
\end{equation}
where $AP(z)^{\text{th}} \equiv \frac{H(z)}{z} \int_{0}^{z} \frac{dz'}{H(z')}$ and $AP(z_i)^{\text{obs}}$ are observational data \cite{Sutter:2012tf,Blake:2011ep,Ross:2006me,Marinoni:2010yoa,daAngela:2005gk,Outram:2003ew,Anderson:2012sa,Paris:2012iw,Schneider:2010hm}.

We are using some data of $H(z)$ of different galaxies from \cite{Simon:2004tf,Stern:2009ep,Moresco:2012jh} and the likelihood function is
\begin{equation}\label{hz}
  \log L_{H(z)} = -\frac{1}{2} \sum_{i=1}^{N}  \left (\frac{H(z_i)^{\text{obs}}-H(z_i)^{\text{th}}}{\sigma_i }\right)^2.
\end{equation}

We consider two cases. In the first case we estimate our model for low $z$ astronomical observations such as SNIa, BAO, $H(z)$ and the Alcock-Paczynski test. The final likelihood function for the first case is
\begin{equation}
 L_{\text{tot}} = L_{\text{SNIa}} L_{\text{BAO}} L_{\text{AP}} L_{H(z)}.
\end{equation}

The second case is for high $z$ and than we use Planck observations, lensing and WP.  The value of $H_0$ is chosen as  $H_0=65.62$ km/(s Mpc).

We use our own code CosmoDarkBox in estimation of the model parameters. The code uses the Metropolis-Hastings algorithm \cite{Metropolis:1953am,Hastings:1970aa} and the dynamical system to obtain the likelihood function \cite{Hu:1995en,Eisenstein:1997ik}.

\begin{figure}[ht]
\centering
\includegraphics[scale=0.4]{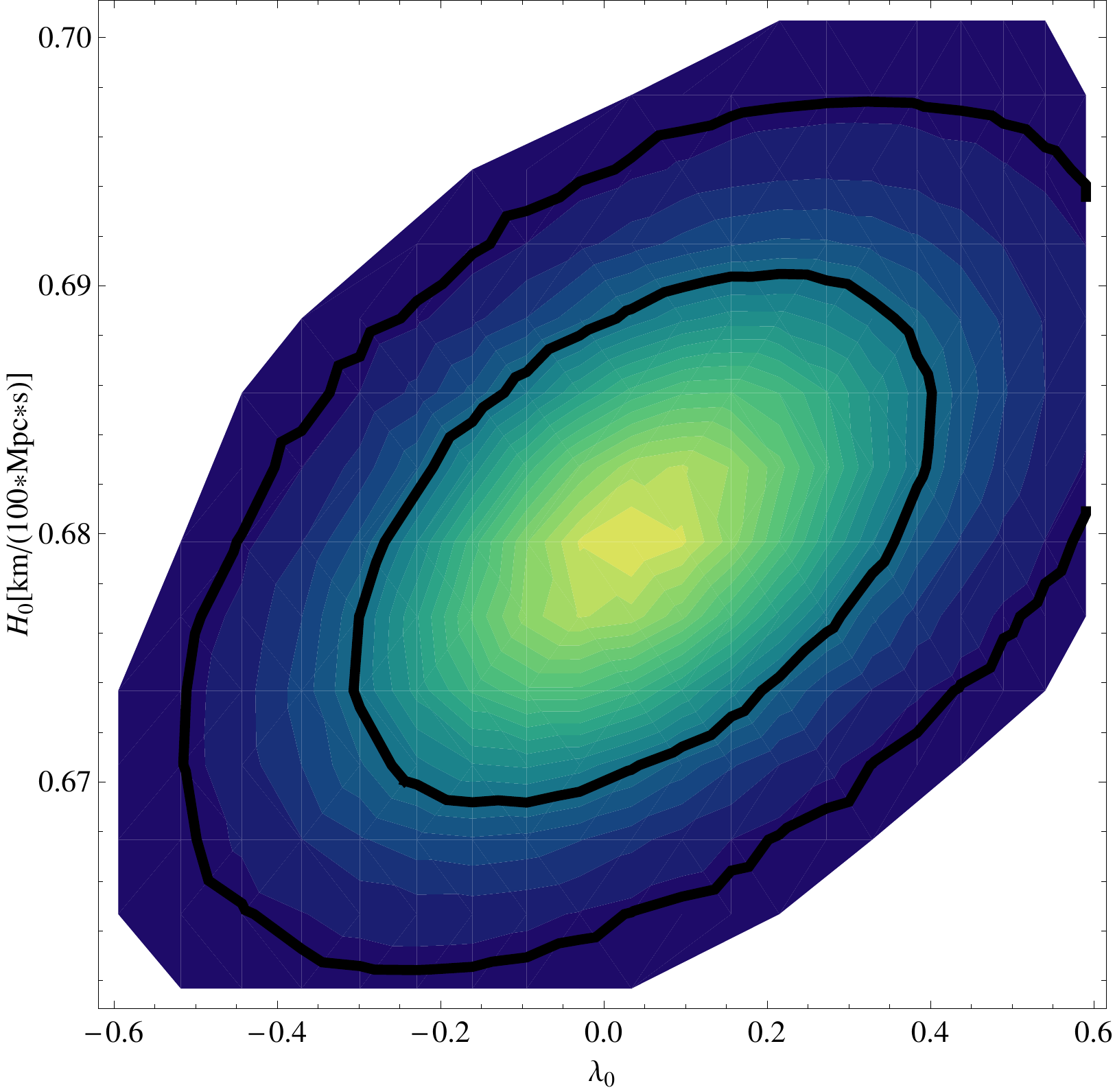}
\caption{The likelihood function of two model parameters ($H_0$, $\lambda_0$) with the marked $68\%$ and $95\%$ confidence levels for SNIa+BAO+$H(z)$+AP test. The value of Hubble constant is estimated from the data as the best fit of value $\Omega_{\text{dm},0} = 0.2650$ and then the diagram of likelihood function is obtained for this value.}
\label{fig:8}
\end{figure}

\begin{figure}[ht]
\centering
\includegraphics[scale=0.4]{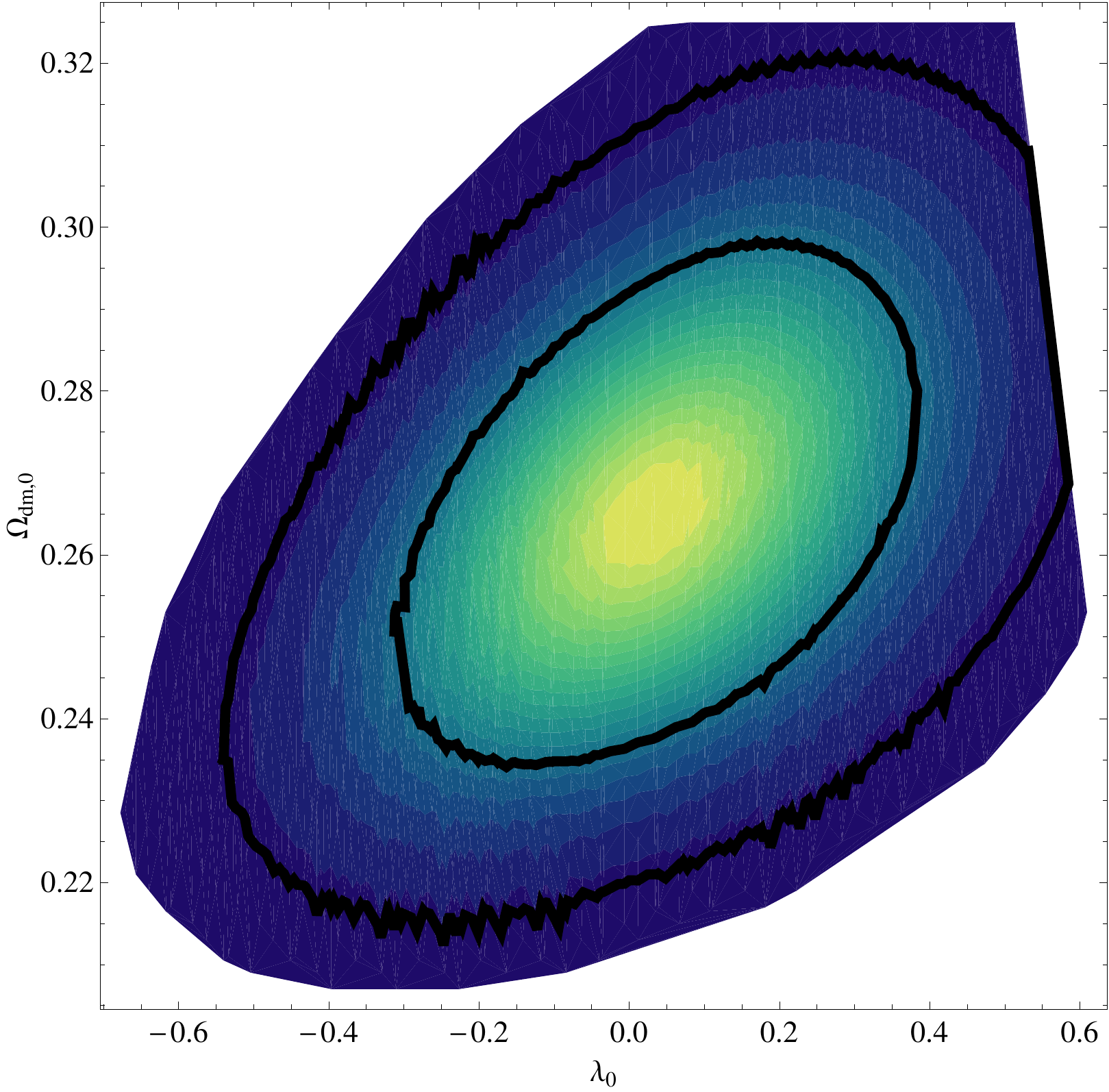}
\caption{The likelihood function of two model parameters ($\Omega_{\text{dm},0}$, $\lambda_0$) with the marked $68\%$ and $95\%$ confidence levels for SNIa+BAO+$H(z)$+AP test. The value of the Hubble constant is estimated from the data as the best fit of value $H_0 = 67.97$ km/(s Mpc) and then the diagram of likelihood function is obtained for this value.}
\label{fig:9}
\end{figure}

\begin{figure}[ht]
\centering
\includegraphics[scale=0.4]{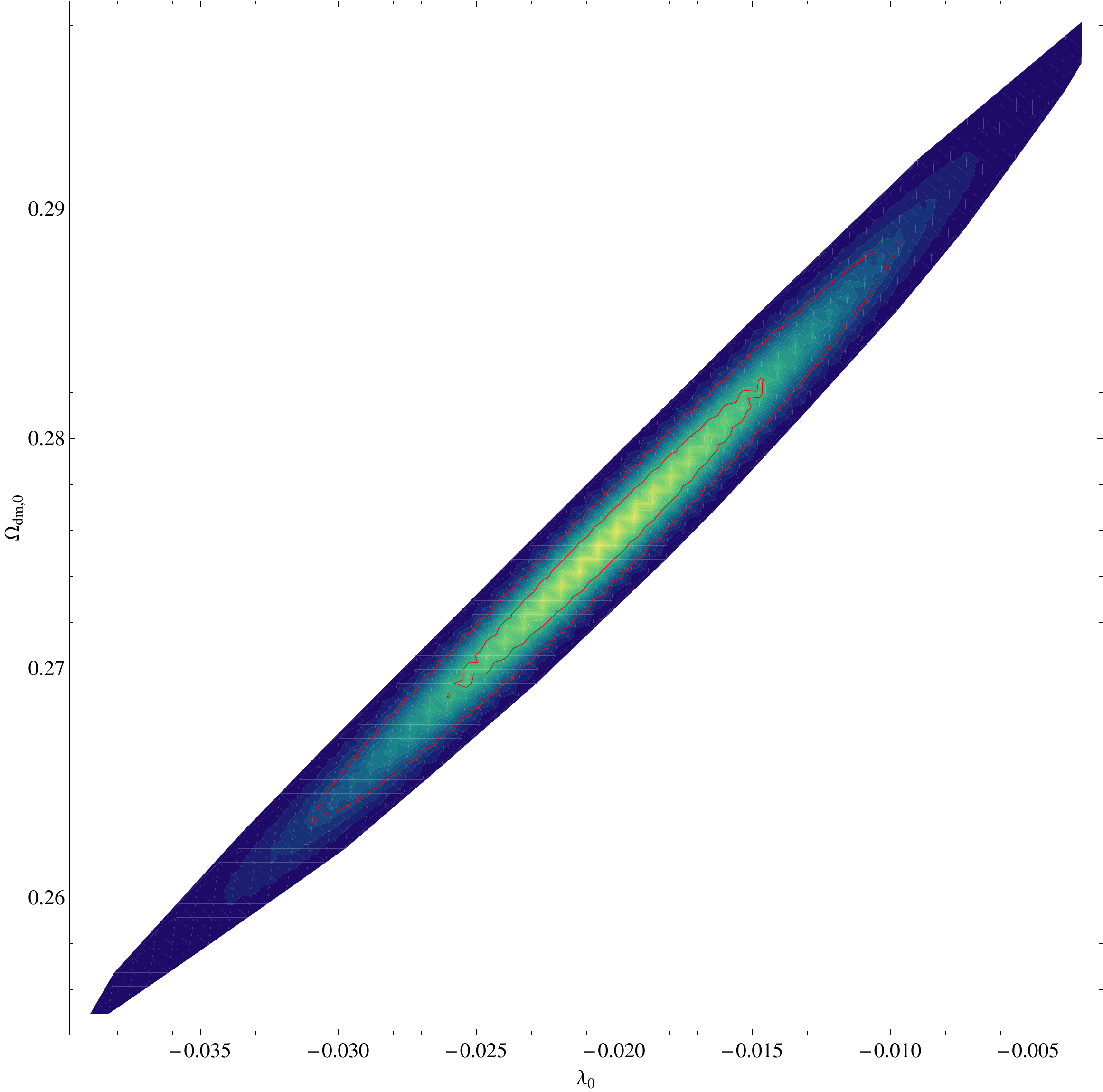}
\caption{The likelihood function of two model parameters ($\Omega_{\text{dm},0}$, $\lambda_0$) with the marked $68\%$ and $95\%$ confidence levels for Planck+lensing+WP. The value of the Hubble constant is equal 65.62 km/(s Mpc) and then the diagram of likelihood function is obtained for this value.}
\label{fig:10}
\end{figure}

The results of statistical analysis for the first case are represented in Table~\ref{table:1}. Figures~\ref{fig:8}, \ref{fig:9} and \ref{fig:10} where it is shown the likelihood function with $68\%$ and $95\%$ confidence level projection on the ($\Omega_{\text{dm},0}$, $\lambda$) plane and the ($H_0$, $\lambda$) plane, respectively. 

For the second case the results of statistical analysis is shown in Table~\ref{table:2}. For the high $z$ data we obtain that $\lambda$ is negative, which means that dark matter particles decay.

\begin{table}
\caption{The best fit and errors for the estimated model for SNIa+BAO+$H(z)$+AP test with $H_0$ from the interval (65.0, 71.0), $\Omega_{\text{dm},0}$ from the interval $(0.20, 0.36)$ and $\lambda_0$ from the interval $(-0.8, 0.8)$. $\Omega_{\text{b}}$ is assumed as 0.048468. The value of $\chi^2$ for the best fit is equal 108.740 and the value of AIC is equal 114.74. In comparison with this model, the $\chi^2$ of the best fit of $\Lambda$CDM is equal 108.761 and AIC is equal 112.761.}
\label{table:1}
\begin{center}
\begin{tabular}{llll} \hline
parameter & best fit & $ 68\% $ CL & $ 95\% $ CL  \\ \hline \hline
$H_0$ & 67.97 km/(s Mpc) & $\begin{array}{c}
   + 1.08\\ -1.06
\end{array}$ & $\begin{array}{c}
   +1.77\\ -1.68
\end{array}$ \\ \hline
$\Omega_{\text{dm},0}$ & 0.2650 & $\begin{array}{c}
   + 0.0333\\ -0.0306
\end{array}$ & $\begin{array}{c}
   +0.0561 \\ -0.0488
\end{array}$\\ \hline
$\lambda$ & 0.0338 & $\begin{array}{c}
   + 0.3832\\ -0.3730
\end{array}$ & $\begin{array}{c}
   + 0.6406\\ -0.6075
\end{array}$ \\ \hline
\end{tabular}
\end{center}
\end{table}

\begin{table}
\caption{The best fit and errors for the estimated model for Planck+lensing+WP for $H_0=65.62$ $\frac{\text{km}}{\text{s Mpc}}$ , with $\Omega_{\text{dm},0}$ from the interval $(0.20, 0.36)$ and $\lambda_0$ from the interval $(-0.8, 0.8)$. $\Omega_{\text{b}}$ is assumed as 0.048468. The value of $\chi^2$ for the best fit is equal 0.001 and the value of AIC is equal 4.001. The small value of $\chi^2$ is a consequence fitting of the two parameter model to the first peak of the spectrum of CMB. In comparison with this model, the value of $\chi^2$ of the best fit of $\Lambda$CDM for $H_0=0.6562$ is equal 13.929 and the value of AIC is 15.929.}
\label{table:2}
\begin{center}
\begin{tabular}{llll} \hline
parameter & best fit & $ 68\% $ CL & $ 95\% $ CL  \\ \hline \hline
$\Omega_{\text{dm},0}$ & 0.2759 & $\begin{array}{c}
   + 0.0068\\ -0.0072
\end{array}$ & $\begin{array}{c}
   +0.0125 \\ -0.0126
\end{array}$\\ \hline
$\lambda$ & -0.0199 & $\begin{array}{c}
   + 0.0057\\ -0.0063
\end{array}$ & $\begin{array}{c}
   + 0.0102\\ -0.0111
\end{array}$ \\ \hline
\end{tabular}
\end{center}
\end{table}

We can use some  information criteria in scientific practice to choose the best model. One of information criteria  is the Akaike information criterion (AIC), which is given by
\begin{equation}
AIC = -2 \log L + 2d,
\end{equation}
where $L$ is the maximum of the likelihood function and $d$ is the number of model parameters. Model which is the best approximation to the truth from the set under consideration has the smallest value of the AIC quantity. It is convenient to evaluate the differences between the AIC quantities computed for the rest of models from our set and the AIC for the best
one. Those differences $\Delta \text{AIC}$
\begin{equation}
\Delta \text{AIC}_{i} =  \text{AIC}_{i} - \text{AIC}_{min}
\end{equation}
are easy to interpret and allow a quick ``strength of evidence'' for the model considered with respect to the best one.

\section{Conclusions} \label{sec:5}

The main goal of the paper was to investigate in details the dynamic of the model with matter and the running cosmological constant term with respect to the cosmological time. It was assumed that baryonic matter satisfies the equation of the state for dust (i.e. is nonrelativistic). We were interested how the running $\Lambda(t)$ influences on the scaling relation for energy density $\rho_{\text{dm}}$. We have found the deviation from standard scaling $a^{-3}$ for this relation. We explained the source of this deviation showing that $\rho_{\text{dm}}$ decreases more rapidly or slowly like $a^{-3+\delta}$ due to the energy transfer from dark matter to dark energy sector or in the opposite direction. The direction of the energy transfer crucially depends on the sign of $\alpha^2$ constant in the model under consideration.

The value of $\alpha^2$ can be theoretically calculated in the quantum formalism developed by Urbanowski and Raczynska \cite{Urbanowski:2013tfa}. In their paper it was proposed a quantum mechanical effect which can be responsible for instability of charge particles which emit X or $\gamma$ rays. The sign of $\alpha^2$ constant is obtained from the analysis of the decay law of unstable states of particles or its survival probability. In these calculations, the crucial role plays the Breit-Wigner distribution function which gives rise to a negative sign of the $\alpha^2$ constant. For typical particles, decaying processes are describing through this distribution function.

From the cosmological point of view it is interesting that fluctuations of energy of instantaneous states of particles of dark matter can be manifested as fluctuations of the velocity of particles \cite{Urbanowski:2013tfa}. In order this effect gives rise to the emission of the electromagnetic radiation from radio up to ultra-high frequencies. In the context of astrophysics important stays information can be obtained from the observation of X rays or $\gamma$ rays. From X ray CCD instruments, dark matter is searched in keV energy for looking for the non-baryonic X ray signature \cite{Sekiya:2015jsa}.

On the other hand the $\alpha^2$ constant is a dimensionless model parameter which value can be estimated from some astronomical data. Our estimations favor the negative $\alpha^2$ constant for high $z$, i.e. it is favored the decaying vacuum of dark matter particles and the radiative nature of the energy transfer to dark energy sector. 

Sterile neutrinos give rise to a negative sign of the $\alpha^2$ constant because they can be described in terms of the Breit-Wigner distribution function \cite{Urbanowski:2013tfa}. The negative sign of the $\alpha^2$ constant offers a new insight into the cosmological constant problem because the running $\Lambda$ is the growing function of the cosmological time with asymptotic $\Lambda_{\text{bare}}$ at $t\rightarrow\infty$. Therefore this problem disappears in a natural way.

In our model the parameter $\alpha^2$ cannot change a sign. On the other hand estimations showed the positive $\alpha^2$ for low redshift data and the negative one for high redshift data. One way of explanation of this discrepancy is to treat the parameter $\alpha^2$ as some effective parameter describing two-component dark matter. For one component the contribution to $\alpha^2$ is negative and for second one is positive. During the cosmic evolution the former decays entirely and the latter survives and there is the switch of the effective $\alpha^2$ sign. 

In our paper we have also found the physical background of the approach developed by Alcaniz and Lima \cite{Alcaniz:2005dg} in which the relation $\rho_{\text{dm}}\propto a^{-3+\lambda}$, where $\lambda=\text{const}$, plays an important role. Our observational analysis of the evolution this parameter during the cosmic evolution indicates that such an ansatz has a strongly physical justification. We showed using astronomical data that the model with decaying dark matter is favored on a 2$\sigma$ level over the $\Lambda$CDM standard cosmological model.

The good news are also coming from our consideration. In interacting cosmology the interacting term which is postulated in different physical forms is interpreted as a kind of nongravitional interactions in the dark sector. We suggest that this interaction has the radiation nature and can be rather interpreted following the Urbanowski and Raczynska idea as a possible emission of cosmic X and $\gamma$ rays by unstable particles \cite{Urbanowski:2013tfa}.

It is still an open discussion about the nature of dark matter: cold or warm dark matter \cite{Lapi:2015zea}. Our results showed that in the model of dark matter decay dark matter particle being lighter than CDM particles. Therefore particles of warm dark matter remain relativistic longer during the cosmic evolution at the early universe.

In the investigating dynamics of the interacting cosmology the corresponding dynamical systems, which are determining the evolutional paths, are not closed until one specify the form of the interacting term $Q$. Usually this form is postulated as a specific function of the Hubble parameter, energy density of matter or scalar field or their time derivatives \cite{Amendola:1999er,Zimdahl:2001ar,Chimento:2009hj,Szydlowski:2005kv}. The case study of our model with decaying dark matter favor the choice of the interacting term in the form $Q \propto H\rho_{\text{m}}$. Zilioti et al. has recently demonstrated that the coincidence problem can be naturally solved in a large class of running $\Lambda(H)$ vacuum cosmologies \cite{Zilioti:2015rza}. The case study of our model fully confirms this idea. 

\subsection*{Acknowledgements}
The work was supported by the grant NCN DEC-2013/09/B/ST2/03455. The authors thank prof. A. Borowiec and A. Krawiec for remarks and comments.


\begin{thebibliography}{49}%
\makeatletter
\providecommand \@ifxundefined [1]{%
 \@ifx{#1\undefined}
}%
\providecommand \@ifnum [1]{%
 \ifnum #1\expandafter \@firstoftwo
 \else \expandafter \@secondoftwo
 \fi
}%
\providecommand \@ifx [1]{%
 \ifx #1\expandafter \@firstoftwo
 \else \expandafter \@secondoftwo
 \fi
}%
\providecommand \natexlab [1]{#1}%
\providecommand \enquote  [1]{``#1''}%
\providecommand \bibnamefont  [1]{#1}%
\providecommand \bibfnamefont [1]{#1}%
\providecommand \citenamefont [1]{#1}%
\providecommand \href@noop [0]{\@secondoftwo}%
\providecommand \href [0]{\begingroup \@sanitize@url \@href}%
\providecommand \@href[1]{\@@startlink{#1}\@@href}%
\providecommand \@@href[1]{\endgroup#1\@@endlink}%
\providecommand \@sanitize@url [0]{\catcode `\\12\catcode `\$12\catcode
  `\&12\catcode `\#12\catcode `\^12\catcode `\_12\catcode `\%12\relax}%
\providecommand \@@startlink[1]{}%
\providecommand \@@endlink[0]{}%
\providecommand \url  [0]{\begingroup\@sanitize@url \@url }%
\providecommand \@url [1]{\endgroup\@href {#1}{\urlprefix }}%
\providecommand \urlprefix  [0]{URL }%
\providecommand \Eprint [0]{\href }%
\providecommand \doibase [0]{http://dx.doi.org/}%
\providecommand \selectlanguage [0]{\@gobble}%
\providecommand \bibinfo  [0]{\@secondoftwo}%
\providecommand \bibfield  [0]{\@secondoftwo}%
\providecommand \translation [1]{[#1]}%
\providecommand \BibitemOpen [0]{}%
\providecommand \bibitemStop [0]{}%
\providecommand \bibitemNoStop [0]{.\EOS\space}%
\providecommand \EOS [0]{\spacefactor3000\relax}%
\providecommand \BibitemShut  [1]{\csname bibitem#1\endcsname}%
\let\auto@bib@innerbib\@empty
\bibitem [{\citenamefont {Szydlowski}\ and\ \citenamefont
  {Stachowski}(2015)}]{Szydlowski:2015rga}%
  \BibitemOpen
  \bibfield  {author} {\bibinfo {author} {\bibfnamefont {M.}~\bibnamefont
  {Szydlowski}}\ and\ \bibinfo {author} {\bibfnamefont {A.}~\bibnamefont
  {Stachowski}},\ }\href@noop {} {\  (\bibinfo {year} {2015})},\ \Eprint
  {http://arxiv.org/abs/1507.02114} {arXiv:1507.02114 [astro-ph.CO]}
  \BibitemShut {NoStop}%
\bibitem [{\citenamefont {Lopez}\ and\ \citenamefont
  {Nanopoulos}(1996)}]{Lopez:1995eb}%
  \BibitemOpen
  \bibfield  {author} {\bibinfo {author} {\bibfnamefont {J.~L.}\ \bibnamefont
  {Lopez}}\ and\ \bibinfo {author} {\bibfnamefont {D.~V.}\ \bibnamefont
  {Nanopoulos}},\ }\href {\doibase 10.1142/S0217732396000023} {\bibfield
  {journal} {\bibinfo  {journal} {Mod. Phys. Lett.}\ }\textbf {\bibinfo
  {volume} {A11}},\ \bibinfo {pages} {1} (\bibinfo {year} {1996})},\ \Eprint
  {http://arxiv.org/abs/hep-ph/9501293} {arXiv:hep-ph/9501293 [hep-ph]}
  \BibitemShut {NoStop}%
\bibitem [{\citenamefont {Wang}\ and\ \citenamefont
  {Meng}(2005)}]{Wang:2004cp}%
  \BibitemOpen
  \bibfield  {author} {\bibinfo {author} {\bibfnamefont {P.}~\bibnamefont
  {Wang}}\ and\ \bibinfo {author} {\bibfnamefont {X.-H.}\ \bibnamefont
  {Meng}},\ }\href {\doibase 10.1088/0264-9381/22/2/003} {\bibfield  {journal}
  {\bibinfo  {journal} {Class. Quant. Grav.}\ }\textbf {\bibinfo {volume}
  {22}},\ \bibinfo {pages} {283} (\bibinfo {year} {2005})},\ \Eprint
  {http://arxiv.org/abs/astro-ph/0408495} {arXiv:astro-ph/0408495 [astro-ph]}
  \BibitemShut {NoStop}%
\bibitem [{\citenamefont {Alcaniz}\ and\ \citenamefont
  {Lima}(2005)}]{Alcaniz:2005dg}%
  \BibitemOpen
  \bibfield  {author} {\bibinfo {author} {\bibfnamefont {J.~S.}\ \bibnamefont
  {Alcaniz}}\ and\ \bibinfo {author} {\bibfnamefont {J.~A.~S.}\ \bibnamefont
  {Lima}},\ }\href {\doibase 10.1103/PhysRevD.72.063516} {\bibfield  {journal}
  {\bibinfo  {journal} {Phys. Rev.}\ }\textbf {\bibinfo {volume} {D72}},\
  \bibinfo {pages} {063516} (\bibinfo {year} {2005})},\ \Eprint
  {http://arxiv.org/abs/astro-ph/0507372} {arXiv:astro-ph/0507372 [astro-ph]}
  \BibitemShut {NoStop}%
\bibitem [{\citenamefont {Graef}\ \emph {et~al.}(2014)\citenamefont {Graef},
  \citenamefont {Costa},\ and\ \citenamefont {Lima}}]{Graef:2013iia}%
  \BibitemOpen
  \bibfield  {author} {\bibinfo {author} {\bibfnamefont {L.~L.}\ \bibnamefont
  {Graef}}, \bibinfo {author} {\bibfnamefont {F.~E.~M.}\ \bibnamefont {Costa}},
  \ and\ \bibinfo {author} {\bibfnamefont {J.~A.~S.}\ \bibnamefont {Lima}},\
  }\href {\doibase 10.1016/j.physletb.2013.12.027} {\bibfield  {journal}
  {\bibinfo  {journal} {Phys. Lett.}\ }\textbf {\bibinfo {volume} {B728}},\
  \bibinfo {pages} {400} (\bibinfo {year} {2014})},\ \Eprint
  {http://arxiv.org/abs/1303.2075} {arXiv:1303.2075 [astro-ph.CO]} \BibitemShut
  {NoStop}%
\bibitem [{\citenamefont {Bessada}\ and\ \citenamefont
  {Miranda}(2013)}]{Bessada:2013maa}%
  \BibitemOpen
  \bibfield  {author} {\bibinfo {author} {\bibfnamefont {D.}~\bibnamefont
  {Bessada}}\ and\ \bibinfo {author} {\bibfnamefont {O.~D.}\ \bibnamefont
  {Miranda}},\ }\href {\doibase 10.1103/PhysRevD.88.083530} {\bibfield
  {journal} {\bibinfo  {journal} {Phys. Rev.}\ }\textbf {\bibinfo {volume}
  {D88}},\ \bibinfo {pages} {083530} (\bibinfo {year} {2013})},\ \Eprint
  {http://arxiv.org/abs/1310.8571} {arXiv:1310.8571 [gr-qc]} \BibitemShut
  {NoStop}%
\bibitem [{\citenamefont {Szydlowski}(2015)}]{Szydlowski:2015bwa}%
  \BibitemOpen
  \bibfield  {author} {\bibinfo {author} {\bibfnamefont {M.}~\bibnamefont
  {Szydlowski}},\ }\href {\doibase 10.1103/PhysRevD.91.123538} {\bibfield
  {journal} {\bibinfo  {journal} {Phys. Rev.}\ }\textbf {\bibinfo {volume}
  {D91}},\ \bibinfo {pages} {123538} (\bibinfo {year} {2015})},\ \Eprint
  {http://arxiv.org/abs/1502.04737} {arXiv:1502.04737 [astro-ph.CO]}
  \BibitemShut {NoStop}%
\bibitem [{\citenamefont {Goncalves}\ \emph {et~al.}(2015)\citenamefont
  {Goncalves}, \citenamefont {Carvalho},\ and\ \citenamefont
  {Alcaniz}}]{Goncalves:2015eaa}%
  \BibitemOpen
  \bibfield  {author} {\bibinfo {author} {\bibfnamefont {R.~S.}\ \bibnamefont
  {Goncalves}}, \bibinfo {author} {\bibfnamefont {G.~C.}\ \bibnamefont
  {Carvalho}}, \ and\ \bibinfo {author} {\bibfnamefont {J.~S.}\ \bibnamefont
  {Alcaniz}},\ }\href@noop {} {\  (\bibinfo {year} {2015})},\ \Eprint
  {http://arxiv.org/abs/1507.01921} {arXiv:1507.01921 [astro-ph.CO]}
  \BibitemShut {NoStop}%
\bibitem [{\citenamefont {del Campo}\ \emph {et~al.}(2015)\citenamefont {del
  Campo}, \citenamefont {Herrera},\ and\ \citenamefont
  {Pavon}}]{delCampo:2015vha}%
  \BibitemOpen
  \bibfield  {author} {\bibinfo {author} {\bibfnamefont {S.}~\bibnamefont {del
  Campo}}, \bibinfo {author} {\bibfnamefont {R.}~\bibnamefont {Herrera}}, \
  and\ \bibinfo {author} {\bibfnamefont {D.}~\bibnamefont {Pavon}},\ }\href
  {\doibase 10.1103/PhysRevD.91.123539} {\bibfield  {journal} {\bibinfo
  {journal} {Phys. Rev.}\ }\textbf {\bibinfo {volume} {D91}},\ \bibinfo {pages}
  {123539} (\bibinfo {year} {2015})},\ \Eprint
  {http://arxiv.org/abs/1507.00187} {arXiv:1507.00187 [gr-qc]} \BibitemShut
  {NoStop}%
\bibitem [{\citenamefont {Urbanowski}\ and\ \citenamefont
  {Raczynska}(2014)}]{Urbanowski:2013tfa}%
  \BibitemOpen
  \bibfield  {author} {\bibinfo {author} {\bibfnamefont {K.}~\bibnamefont
  {Urbanowski}}\ and\ \bibinfo {author} {\bibfnamefont {K.}~\bibnamefont
  {Raczynska}},\ }\href {\doibase 10.1016/j.physletb.2014.02.043} {\bibfield
  {journal} {\bibinfo  {journal} {Phys. Lett.}\ }\textbf {\bibinfo {volume}
  {B731}},\ \bibinfo {pages} {236} (\bibinfo {year} {2014})},\ \Eprint
  {http://arxiv.org/abs/1303.6975} {arXiv:1303.6975 [astro-ph.HE]} \BibitemShut
  {NoStop}%
\bibitem [{\citenamefont {Minazzoli}\ and\ \citenamefont
  {Harko}(2012)}]{Minazzoli:2012md}%
  \BibitemOpen
  \bibfield  {author} {\bibinfo {author} {\bibfnamefont {O.}~\bibnamefont
  {Minazzoli}}\ and\ \bibinfo {author} {\bibfnamefont {T.}~\bibnamefont
  {Harko}},\ }\href {\doibase 10.1103/PhysRevD.86.087502} {\bibfield  {journal}
  {\bibinfo  {journal} {Phys. Rev.}\ }\textbf {\bibinfo {volume} {D86}},\
  \bibinfo {pages} {087502} (\bibinfo {year} {2012})},\ \Eprint
  {http://arxiv.org/abs/1209.2754} {arXiv:1209.2754 [gr-qc]} \BibitemShut
  {NoStop}%
\bibitem [{\citenamefont {Boyarsky}\ \emph {et~al.}(2009)\citenamefont
  {Boyarsky}, \citenamefont {Ruchayskiy},\ and\ \citenamefont
  {Shaposhnikov}}]{Boyarsky:2009ix}%
  \BibitemOpen
  \bibfield  {author} {\bibinfo {author} {\bibfnamefont {A.}~\bibnamefont
  {Boyarsky}}, \bibinfo {author} {\bibfnamefont {O.}~\bibnamefont
  {Ruchayskiy}}, \ and\ \bibinfo {author} {\bibfnamefont {M.}~\bibnamefont
  {Shaposhnikov}},\ }\href {\doibase 10.1146/annurev.nucl.010909.083654}
  {\bibfield  {journal} {\bibinfo  {journal} {Ann. Rev. Nucl. Part. Sci.}\
  }\textbf {\bibinfo {volume} {59}},\ \bibinfo {pages} {191} (\bibinfo {year}
  {2009})},\ \Eprint {http://arxiv.org/abs/0901.0011} {arXiv:0901.0011
  [hep-ph]} \BibitemShut {NoStop}%
\bibitem [{\citenamefont {Motohashi}\ \emph {et~al.}(2013)\citenamefont
  {Motohashi}, \citenamefont {Starobinsky},\ and\ \citenamefont
  {Yokoyama}}]{Motohashi:2012wc}%
  \BibitemOpen
  \bibfield  {author} {\bibinfo {author} {\bibfnamefont {H.}~\bibnamefont
  {Motohashi}}, \bibinfo {author} {\bibfnamefont {A.~A.}\ \bibnamefont
  {Starobinsky}}, \ and\ \bibinfo {author} {\bibfnamefont {J.}~\bibnamefont
  {Yokoyama}},\ }\href {\doibase 10.1103/PhysRevLett.110.121302} {\bibfield
  {journal} {\bibinfo  {journal} {Phys. Rev. Lett.}\ }\textbf {\bibinfo
  {volume} {110}},\ \bibinfo {pages} {121302} (\bibinfo {year} {2013})},\
  \Eprint {http://arxiv.org/abs/1203.6828} {arXiv:1203.6828 [astro-ph.CO]}
  \BibitemShut {NoStop}%
\bibitem [{\citenamefont {Abazajian}\ \emph {et~al.}(2001)\citenamefont
  {Abazajian}, \citenamefont {Fuller},\ and\ \citenamefont
  {Tucker}}]{Abazajian:2001vt}%
  \BibitemOpen
  \bibfield  {author} {\bibinfo {author} {\bibfnamefont {K.}~\bibnamefont
  {Abazajian}}, \bibinfo {author} {\bibfnamefont {G.~M.}\ \bibnamefont
  {Fuller}}, \ and\ \bibinfo {author} {\bibfnamefont {W.~H.}\ \bibnamefont
  {Tucker}},\ }\href {\doibase 10.1086/323867} {\bibfield  {journal} {\bibinfo
  {journal} {Astrophys. J.}\ }\textbf {\bibinfo {volume} {562}},\ \bibinfo
  {pages} {593} (\bibinfo {year} {2001})},\ \Eprint
  {http://arxiv.org/abs/astro-ph/0106002} {arXiv:astro-ph/0106002 [astro-ph]}
  \BibitemShut {NoStop}%
\bibitem [{\citenamefont {Boyarsky}\ \emph {et~al.}(2014)\citenamefont
  {Boyarsky}, \citenamefont {Ruchayskiy}, \citenamefont {Iakubovskyi},\ and\
  \citenamefont {Franse}}]{Boyarsky:2014jta}%
  \BibitemOpen
  \bibfield  {author} {\bibinfo {author} {\bibfnamefont {A.}~\bibnamefont
  {Boyarsky}}, \bibinfo {author} {\bibfnamefont {O.}~\bibnamefont
  {Ruchayskiy}}, \bibinfo {author} {\bibfnamefont {D.}~\bibnamefont
  {Iakubovskyi}}, \ and\ \bibinfo {author} {\bibfnamefont {J.}~\bibnamefont
  {Franse}},\ }\href {\doibase 10.1103/PhysRevLett.113.251301} {\bibfield
  {journal} {\bibinfo  {journal} {Phys. Rev. Lett.}\ }\textbf {\bibinfo
  {volume} {113}},\ \bibinfo {pages} {251301} (\bibinfo {year} {2014})},\
  \Eprint {http://arxiv.org/abs/1402.4119} {arXiv:1402.4119 [astro-ph.CO]}
  \BibitemShut {NoStop}%
\bibitem [{\citenamefont {Dolgov}\ and\ \citenamefont
  {Hansen}(2002)}]{Dolgov:2000ew}%
  \BibitemOpen
  \bibfield  {author} {\bibinfo {author} {\bibfnamefont {A.~D.}\ \bibnamefont
  {Dolgov}}\ and\ \bibinfo {author} {\bibfnamefont {S.~H.}\ \bibnamefont
  {Hansen}},\ }\href {\doibase 10.1016/S0927-6505(01)00115-3} {\bibfield
  {journal} {\bibinfo  {journal} {Astropart. Phys.}\ }\textbf {\bibinfo
  {volume} {16}},\ \bibinfo {pages} {339} (\bibinfo {year} {2002})},\ \Eprint
  {http://arxiv.org/abs/hep-ph/0009083} {arXiv:hep-ph/0009083 [hep-ph]}
  \BibitemShut {NoStop}%
\bibitem [{\citenamefont {Sekiya}\ \emph {et~al.}(2015)\citenamefont {Sekiya},
  \citenamefont {Yamasaki},\ and\ \citenamefont {Mitsuda}}]{Sekiya:2015jsa}%
  \BibitemOpen
  \bibfield  {author} {\bibinfo {author} {\bibfnamefont {N.}~\bibnamefont
  {Sekiya}}, \bibinfo {author} {\bibfnamefont {N.~Y.}\ \bibnamefont
  {Yamasaki}}, \ and\ \bibinfo {author} {\bibfnamefont {K.}~\bibnamefont
  {Mitsuda}},\ }\href@noop {} {\  (\bibinfo {year} {2015})},\ \Eprint
  {http://arxiv.org/abs/1504.02826} {arXiv:1504.02826 [astro-ph.HE]}
  \BibitemShut {NoStop}%
\bibitem [{\citenamefont {Amendola}\ \emph {et~al.}(2007)\citenamefont
  {Amendola}, \citenamefont {Camargo~Campos},\ and\ \citenamefont
  {Rosenfeld}}]{Amendola:2006dg}%
  \BibitemOpen
  \bibfield  {author} {\bibinfo {author} {\bibfnamefont {L.}~\bibnamefont
  {Amendola}}, \bibinfo {author} {\bibfnamefont {G.}~\bibnamefont
  {Camargo~Campos}}, \ and\ \bibinfo {author} {\bibfnamefont {R.}~\bibnamefont
  {Rosenfeld}},\ }\href {\doibase 10.1103/PhysRevD.75.083506} {\bibfield
  {journal} {\bibinfo  {journal} {Phys. Rev.}\ }\textbf {\bibinfo {volume}
  {D75}},\ \bibinfo {pages} {083506} (\bibinfo {year} {2007})},\ \Eprint
  {http://arxiv.org/abs/astro-ph/0610806} {arXiv:astro-ph/0610806 [astro-ph]}
  \BibitemShut {NoStop}%
\bibitem [{\citenamefont {Majerotto}\ \emph {et~al.}(2004)\citenamefont
  {Majerotto}, \citenamefont {Sapone},\ and\ \citenamefont
  {Amendola}}]{Majerotto:2004ji}%
  \BibitemOpen
  \bibfield  {author} {\bibinfo {author} {\bibfnamefont {E.}~\bibnamefont
  {Majerotto}}, \bibinfo {author} {\bibfnamefont {D.}~\bibnamefont {Sapone}}, \
  and\ \bibinfo {author} {\bibfnamefont {L.}~\bibnamefont {Amendola}},\
  }\href@noop {} {\  (\bibinfo {year} {2004})},\ \Eprint
  {http://arxiv.org/abs/astro-ph/0410543} {arXiv:astro-ph/0410543 [astro-ph]}
  \BibitemShut {NoStop}%
\bibitem [{\citenamefont {Rosenfeld}(2005)}]{Rosenfeld:2005pw}%
  \BibitemOpen
  \bibfield  {author} {\bibinfo {author} {\bibfnamefont {R.}~\bibnamefont
  {Rosenfeld}},\ }\href {\doibase 10.1016/j.physletb.2005.08.049} {\bibfield
  {journal} {\bibinfo  {journal} {Phys. Lett.}\ }\textbf {\bibinfo {volume}
  {B624}},\ \bibinfo {pages} {158} (\bibinfo {year} {2005})},\ \Eprint
  {http://arxiv.org/abs/astro-ph/0504121} {arXiv:astro-ph/0504121 [astro-ph]}
  \BibitemShut {NoStop}%
\bibitem [{\citenamefont {Suzuki}\ \emph {et~al.}(2012)\citenamefont {Suzuki}
  \emph {et~al.}}]{Suzuki:2011hu}%
  \BibitemOpen
  \bibfield  {author} {\bibinfo {author} {\bibfnamefont {N.}~\bibnamefont
  {Suzuki}} \emph {et~al.},\ }\href {\doibase 10.1088/0004-637X/746/1/85}
  {\bibfield  {journal} {\bibinfo  {journal} {Astrophys. J.}\ }\textbf
  {\bibinfo {volume} {746}},\ \bibinfo {pages} {85} (\bibinfo {year} {2012})},\
  \Eprint {http://arxiv.org/abs/1105.3470} {arXiv:1105.3470 [astro-ph.CO]}
  \BibitemShut {NoStop}%
\bibitem [{\citenamefont {Percival}\ \emph {et~al.}(2010)\citenamefont
  {Percival} \emph {et~al.}}]{Percival:2009xn}%
  \BibitemOpen
  \bibfield  {author} {\bibinfo {author} {\bibfnamefont {W.~J.}\ \bibnamefont
  {Percival}} \emph {et~al.} (\bibinfo {collaboration} {SDSS}),\ }\href
  {\doibase 10.1111/j.1365-2966.2009.15812.x} {\bibfield  {journal} {\bibinfo
  {journal} {Mon. Not. Roy. Astron. Soc.}\ }\textbf {\bibinfo {volume} {401}},\
  \bibinfo {pages} {2148} (\bibinfo {year} {2010})},\ \Eprint
  {http://arxiv.org/abs/0907.1660} {arXiv:0907.1660 [astro-ph.CO]} \BibitemShut
  {NoStop}%
\bibitem [{\citenamefont {Beutler}\ \emph {et~al.}(2011)\citenamefont
  {Beutler}, \citenamefont {Blake}, \citenamefont {Colless}, \citenamefont
  {Jones}, \citenamefont {Staveley-Smith}, \citenamefont {Campbell},
  \citenamefont {Parker}, \citenamefont {Saunders},\ and\ \citenamefont
  {Watson}}]{Beutler:2011hx}%
  \BibitemOpen
  \bibfield  {author} {\bibinfo {author} {\bibfnamefont {F.}~\bibnamefont
  {Beutler}}, \bibinfo {author} {\bibfnamefont {C.}~\bibnamefont {Blake}},
  \bibinfo {author} {\bibfnamefont {M.}~\bibnamefont {Colless}}, \bibinfo
  {author} {\bibfnamefont {D.~H.}\ \bibnamefont {Jones}}, \bibinfo {author}
  {\bibfnamefont {L.}~\bibnamefont {Staveley-Smith}}, \bibinfo {author}
  {\bibfnamefont {L.}~\bibnamefont {Campbell}}, \bibinfo {author}
  {\bibfnamefont {Q.}~\bibnamefont {Parker}}, \bibinfo {author} {\bibfnamefont
  {W.}~\bibnamefont {Saunders}}, \ and\ \bibinfo {author} {\bibfnamefont
  {F.}~\bibnamefont {Watson}},\ }\href {\doibase
  10.1111/j.1365-2966.2011.19250.x} {\bibfield  {journal} {\bibinfo  {journal}
  {Mon. Not. Roy. Astron. Soc.}\ }\textbf {\bibinfo {volume} {416}},\ \bibinfo
  {pages} {3017} (\bibinfo {year} {2011})},\ \Eprint
  {http://arxiv.org/abs/1106.3366} {arXiv:1106.3366 [astro-ph.CO]} \BibitemShut
  {NoStop}%
\bibitem [{\citenamefont {Anderson}\ \emph {et~al.}(2013)\citenamefont
  {Anderson} \emph {et~al.}}]{Anderson:2012sa}%
  \BibitemOpen
  \bibfield  {author} {\bibinfo {author} {\bibfnamefont {L.}~\bibnamefont
  {Anderson}} \emph {et~al.},\ }\href {\doibase
  10.1111/j.1365-2966.2012.22066.x} {\bibfield  {journal} {\bibinfo  {journal}
  {Mon. Not. Roy. Astron. Soc.}\ }\textbf {\bibinfo {volume} {427}},\ \bibinfo
  {pages} {3435} (\bibinfo {year} {2013})},\ \Eprint
  {http://arxiv.org/abs/1203.6594} {arXiv:1203.6594 [astro-ph.CO]} \BibitemShut
  {NoStop}%
\bibitem [{\citenamefont {Blake}\ \emph {et~al.}(2012)\citenamefont {Blake}
  \emph {et~al.}}]{Blake:2012pj}%
  \BibitemOpen
  \bibfield  {author} {\bibinfo {author} {\bibfnamefont {C.}~\bibnamefont
  {Blake}} \emph {et~al.},\ }\href {\doibase 10.1111/j.1365-2966.2012.21473.x}
  {\bibfield  {journal} {\bibinfo  {journal} {Mon. Not. Roy. Astron. Soc.}\
  }\textbf {\bibinfo {volume} {425}},\ \bibinfo {pages} {405} (\bibinfo {year}
  {2012})},\ \Eprint {http://arxiv.org/abs/1204.3674} {arXiv:1204.3674
  [astro-ph.CO]} \BibitemShut {NoStop}%
\bibitem [{\citenamefont {Eisenstein}\ and\ \citenamefont
  {Hu}(1998)}]{Eisenstein:1997ik}%
  \BibitemOpen
  \bibfield  {author} {\bibinfo {author} {\bibfnamefont {D.~J.}\ \bibnamefont
  {Eisenstein}}\ and\ \bibinfo {author} {\bibfnamefont {W.}~\bibnamefont
  {Hu}},\ }\href {\doibase 10.1086/305424} {\bibfield  {journal} {\bibinfo
  {journal} {Astrophys. J.}\ }\textbf {\bibinfo {volume} {496}},\ \bibinfo
  {pages} {605} (\bibinfo {year} {1998})},\ \Eprint
  {http://arxiv.org/abs/astro-ph/9709112} {arXiv:astro-ph/9709112 [astro-ph]}
  \BibitemShut {NoStop}%
\bibitem [{\citenamefont {Ade}\ \emph {et~al.}(2014)\citenamefont {Ade} \emph
  {et~al.}}]{Ade:2013zuv}%
  \BibitemOpen
  \bibfield  {author} {\bibinfo {author} {\bibfnamefont {P.~A.~R.}\
  \bibnamefont {Ade}} \emph {et~al.} (\bibinfo {collaboration} {Planck}),\
  }\href {\doibase 10.1051/0004-6361/201321591} {\bibfield  {journal} {\bibinfo
   {journal} {Astron. Astrophys.}\ }\textbf {\bibinfo {volume} {571}},\
  \bibinfo {pages} {A16} (\bibinfo {year} {2014})},\ \Eprint
  {http://arxiv.org/abs/1303.5076} {arXiv:1303.5076 [astro-ph.CO]} \BibitemShut
  {NoStop}%
\bibitem [{\citenamefont {Alcock}\ and\ \citenamefont
  {Paczynski}(1979)}]{Alcock:1979mp}%
  \BibitemOpen
  \bibfield  {author} {\bibinfo {author} {\bibfnamefont {C.}~\bibnamefont
  {Alcock}}\ and\ \bibinfo {author} {\bibfnamefont {B.}~\bibnamefont
  {Paczynski}},\ }\href {\doibase 10.1038/281358a0} {\bibfield  {journal}
  {\bibinfo  {journal} {Nature}\ }\textbf {\bibinfo {volume} {281}},\ \bibinfo
  {pages} {358} (\bibinfo {year} {1979})}\BibitemShut {NoStop}%
\bibitem [{\citenamefont {Lopez-Corredoira}(2014)}]{Lopez-Corredoira:2013lca}%
  \BibitemOpen
  \bibfield  {author} {\bibinfo {author} {\bibfnamefont {M.}~\bibnamefont
  {Lopez-Corredoira}},\ }\href {\doibase 10.1088/0004-637X/781/2/96} {\bibfield
   {journal} {\bibinfo  {journal} {Astrophys. J.}\ }\textbf {\bibinfo {volume}
  {781}},\ \bibinfo {pages} {96} (\bibinfo {year} {2014})},\ \Eprint
  {http://arxiv.org/abs/1312.0003} {arXiv:1312.0003 [astro-ph.CO]} \BibitemShut
  {NoStop}%
\bibitem [{\citenamefont {Sutter}\ \emph {et~al.}(2012)\citenamefont {Sutter},
  \citenamefont {Lavaux}, \citenamefont {Wandelt},\ and\ \citenamefont
  {Weinberg}}]{Sutter:2012tf}%
  \BibitemOpen
  \bibfield  {author} {\bibinfo {author} {\bibfnamefont {P.~M.}\ \bibnamefont
  {Sutter}}, \bibinfo {author} {\bibfnamefont {G.}~\bibnamefont {Lavaux}},
  \bibinfo {author} {\bibfnamefont {B.~D.}\ \bibnamefont {Wandelt}}, \ and\
  \bibinfo {author} {\bibfnamefont {D.~H.}\ \bibnamefont {Weinberg}},\ }\href
  {\doibase 10.1088/0004-637X/761/2/187} {\bibfield  {journal} {\bibinfo
  {journal} {Astrophys. J.}\ }\textbf {\bibinfo {volume} {761}},\ \bibinfo
  {pages} {187} (\bibinfo {year} {2012})},\ \Eprint
  {http://arxiv.org/abs/1208.1058} {arXiv:1208.1058 [astro-ph.CO]} \BibitemShut
  {NoStop}%
\bibitem [{\citenamefont {Blake}\ \emph {et~al.}(2011)\citenamefont {Blake}
  \emph {et~al.}}]{Blake:2011ep}%
  \BibitemOpen
  \bibfield  {author} {\bibinfo {author} {\bibfnamefont {C.}~\bibnamefont
  {Blake}} \emph {et~al.},\ }\href {\doibase 10.1111/j.1365-2966.2011.19606.x}
  {\bibfield  {journal} {\bibinfo  {journal} {Mon. Not. Roy. Astron. Soc.}\
  }\textbf {\bibinfo {volume} {418}},\ \bibinfo {pages} {1725} (\bibinfo {year}
  {2011})},\ \Eprint {http://arxiv.org/abs/1108.2637} {arXiv:1108.2637
  [astro-ph.CO]} \BibitemShut {NoStop}%
\bibitem [{\citenamefont {Ross}\ \emph {et~al.}(2007)\citenamefont {Ross} \emph
  {et~al.}}]{Ross:2006me}%
  \BibitemOpen
  \bibfield  {author} {\bibinfo {author} {\bibfnamefont {N.~P.}\ \bibnamefont
  {Ross}} \emph {et~al.},\ }\href {\doibase 10.1111/j.1365-2966.2007.12289.x}
  {\bibfield  {journal} {\bibinfo  {journal} {Mon. Not. Roy. Astron. Soc.}\
  }\textbf {\bibinfo {volume} {381}},\ \bibinfo {pages} {573} (\bibinfo {year}
  {2007})},\ \Eprint {http://arxiv.org/abs/astro-ph/0612400}
  {arXiv:astro-ph/0612400 [astro-ph]} \BibitemShut {NoStop}%
\bibitem [{\citenamefont {Marinoni}\ and\ \citenamefont
  {Buzzi}(2010)}]{Marinoni:2010yoa}%
  \BibitemOpen
  \bibfield  {author} {\bibinfo {author} {\bibfnamefont {C.}~\bibnamefont
  {Marinoni}}\ and\ \bibinfo {author} {\bibfnamefont {A.}~\bibnamefont
  {Buzzi}},\ }\href {\doibase 10.1038/nature09577} {\bibfield  {journal}
  {\bibinfo  {journal} {Nature}\ }\textbf {\bibinfo {volume} {468}},\ \bibinfo
  {pages} {539} (\bibinfo {year} {2010})}\BibitemShut {NoStop}%
\bibitem [{\citenamefont {da~Angela}\ \emph {et~al.}(2005)\citenamefont
  {da~Angela}, \citenamefont {Outram},\ and\ \citenamefont
  {Shanks}}]{daAngela:2005gk}%
  \BibitemOpen
  \bibfield  {author} {\bibinfo {author} {\bibfnamefont {J.}~\bibnamefont
  {da~Angela}}, \bibinfo {author} {\bibfnamefont {P.~J.}\ \bibnamefont
  {Outram}}, \ and\ \bibinfo {author} {\bibfnamefont {T.}~\bibnamefont
  {Shanks}},\ }\href {\doibase 10.1111/j.1365-2966.2005.09212.x} {\bibfield
  {journal} {\bibinfo  {journal} {Mon. Not. Roy. Astron. Soc.}\ }\textbf
  {\bibinfo {volume} {361}},\ \bibinfo {pages} {879} (\bibinfo {year}
  {2005})},\ \Eprint {http://arxiv.org/abs/astro-ph/0505469}
  {arXiv:astro-ph/0505469 [astro-ph]} \BibitemShut {NoStop}%
\bibitem [{\citenamefont {Outram}\ \emph {et~al.}(2004)\citenamefont {Outram},
  \citenamefont {Shanks}, \citenamefont {Boyle}, \citenamefont {Croom},
  \citenamefont {Hoyle}, \citenamefont {Loaring}, \citenamefont {Miller},\ and\
  \citenamefont {Smith}}]{Outram:2003ew}%
  \BibitemOpen
  \bibfield  {author} {\bibinfo {author} {\bibfnamefont {P.~J.}\ \bibnamefont
  {Outram}}, \bibinfo {author} {\bibfnamefont {T.}~\bibnamefont {Shanks}},
  \bibinfo {author} {\bibfnamefont {B.~J.}\ \bibnamefont {Boyle}}, \bibinfo
  {author} {\bibfnamefont {S.~M.}\ \bibnamefont {Croom}}, \bibinfo {author}
  {\bibfnamefont {F.}~\bibnamefont {Hoyle}}, \bibinfo {author} {\bibfnamefont
  {N.~S.}\ \bibnamefont {Loaring}}, \bibinfo {author} {\bibfnamefont
  {L.}~\bibnamefont {Miller}}, \ and\ \bibinfo {author} {\bibfnamefont {R.~J.}\
  \bibnamefont {Smith}},\ }\href {\doibase 10.1111/j.1365-2966.2004.07348.x}
  {\bibfield  {journal} {\bibinfo  {journal} {Mon. Not. Roy. Astron. Soc.}\
  }\textbf {\bibinfo {volume} {348}},\ \bibinfo {pages} {745} (\bibinfo {year}
  {2004})},\ \Eprint {http://arxiv.org/abs/astro-ph/0310873}
  {arXiv:astro-ph/0310873 [astro-ph]} \BibitemShut {NoStop}%
\bibitem [{\citenamefont {Paris}\ \emph {et~al.}(2012)\citenamefont {Paris}
  \emph {et~al.}}]{Paris:2012iw}%
  \BibitemOpen
  \bibfield  {author} {\bibinfo {author} {\bibfnamefont {I.}~\bibnamefont
  {Paris}} \emph {et~al.},\ }\href {\doibase 10.1051/0004-6361/201220142}
  {\bibfield  {journal} {\bibinfo  {journal} {Astron. Astrophys.}\ }\textbf
  {\bibinfo {volume} {548}},\ \bibinfo {pages} {A66} (\bibinfo {year}
  {2012})},\ \Eprint {http://arxiv.org/abs/1210.5166} {arXiv:1210.5166
  [astro-ph.CO]} \BibitemShut {NoStop}%
\bibitem [{\citenamefont {Schneider}\ \emph {et~al.}(2010)\citenamefont
  {Schneider} \emph {et~al.}}]{Schneider:2010hm}%
  \BibitemOpen
  \bibfield  {author} {\bibinfo {author} {\bibfnamefont {D.~P.}\ \bibnamefont
  {Schneider}} \emph {et~al.} (\bibinfo {collaboration} {SDSS}),\ }\href
  {\doibase 10.1088/0004-6256/139/6/2360} {\bibfield  {journal} {\bibinfo
  {journal} {Astron. J.}\ }\textbf {\bibinfo {volume} {139}},\ \bibinfo {pages}
  {2360} (\bibinfo {year} {2010})},\ \Eprint {http://arxiv.org/abs/1004.1167}
  {arXiv:1004.1167 [astro-ph.CO]} \BibitemShut {NoStop}%
\bibitem [{\citenamefont {Simon}\ \emph {et~al.}(2005)\citenamefont {Simon},
  \citenamefont {Verde},\ and\ \citenamefont {Jimenez}}]{Simon:2004tf}%
  \BibitemOpen
  \bibfield  {author} {\bibinfo {author} {\bibfnamefont {J.}~\bibnamefont
  {Simon}}, \bibinfo {author} {\bibfnamefont {L.}~\bibnamefont {Verde}}, \ and\
  \bibinfo {author} {\bibfnamefont {R.}~\bibnamefont {Jimenez}},\ }\href
  {\doibase 10.1103/PhysRevD.71.123001} {\bibfield  {journal} {\bibinfo
  {journal} {Phys. Rev.}\ }\textbf {\bibinfo {volume} {D71}},\ \bibinfo {pages}
  {123001} (\bibinfo {year} {2005})},\ \Eprint
  {http://arxiv.org/abs/astro-ph/0412269} {arXiv:astro-ph/0412269 [astro-ph]}
  \BibitemShut {NoStop}%
\bibitem [{\citenamefont {Stern}\ \emph {et~al.}(2010)\citenamefont {Stern},
  \citenamefont {Jimenez}, \citenamefont {Verde}, \citenamefont
  {Kamionkowski},\ and\ \citenamefont {Stanford}}]{Stern:2009ep}%
  \BibitemOpen
  \bibfield  {author} {\bibinfo {author} {\bibfnamefont {D.}~\bibnamefont
  {Stern}}, \bibinfo {author} {\bibfnamefont {R.}~\bibnamefont {Jimenez}},
  \bibinfo {author} {\bibfnamefont {L.}~\bibnamefont {Verde}}, \bibinfo
  {author} {\bibfnamefont {M.}~\bibnamefont {Kamionkowski}}, \ and\ \bibinfo
  {author} {\bibfnamefont {S.~A.}\ \bibnamefont {Stanford}},\ }\href {\doibase
  10.1088/1475-7516/2010/02/008} {\bibfield  {journal} {\bibinfo  {journal}
  {JCAP}\ }\textbf {\bibinfo {volume} {1002}},\ \bibinfo {pages} {008}
  (\bibinfo {year} {2010})},\ \Eprint {http://arxiv.org/abs/0907.3149}
  {arXiv:0907.3149 [astro-ph.CO]} \BibitemShut {NoStop}%
\bibitem [{\citenamefont {Moresco}\ \emph {et~al.}(2012)\citenamefont {Moresco}
  \emph {et~al.}}]{Moresco:2012jh}%
  \BibitemOpen
  \bibfield  {author} {\bibinfo {author} {\bibfnamefont {M.}~\bibnamefont
  {Moresco}} \emph {et~al.},\ }\href {\doibase 10.1088/1475-7516/2012/08/006}
  {\bibfield  {journal} {\bibinfo  {journal} {JCAP}\ }\textbf {\bibinfo
  {volume} {1208}},\ \bibinfo {pages} {006} (\bibinfo {year} {2012})},\ \Eprint
  {http://arxiv.org/abs/1201.3609} {arXiv:1201.3609 [astro-ph.CO]} \BibitemShut
  {NoStop}%
\bibitem [{\citenamefont {Metropolis}\ \emph {et~al.}(1953)\citenamefont
  {Metropolis}, \citenamefont {Rosenbluth}, \citenamefont {Rosenbluth},
  \citenamefont {Teller},\ and\ \citenamefont {Teller}}]{Metropolis:1953am}%
  \BibitemOpen
  \bibfield  {author} {\bibinfo {author} {\bibfnamefont {N.}~\bibnamefont
  {Metropolis}}, \bibinfo {author} {\bibfnamefont {A.~W.}\ \bibnamefont
  {Rosenbluth}}, \bibinfo {author} {\bibfnamefont {M.~N.}\ \bibnamefont
  {Rosenbluth}}, \bibinfo {author} {\bibfnamefont {A.~H.}\ \bibnamefont
  {Teller}}, \ and\ \bibinfo {author} {\bibfnamefont {E.}~\bibnamefont
  {Teller}},\ }\href {\doibase 10.1063/1.1699114} {\bibfield  {journal}
  {\bibinfo  {journal} {J. Chem. Phys.}\ }\textbf {\bibinfo {volume} {21}},\
  \bibinfo {pages} {1087} (\bibinfo {year} {1953})}\BibitemShut {NoStop}%
\bibitem [{\citenamefont {Hastings}(1970)}]{Hastings:1970aa}%
  \BibitemOpen
  \bibfield  {author} {\bibinfo {author} {\bibfnamefont {W.~K.}\ \bibnamefont
  {Hastings}},\ }\href {\doibase 10.1093/biomet/57.1.97} {\bibfield  {journal}
  {\bibinfo  {journal} {Biometrika}\ }\textbf {\bibinfo {volume} {57}},\
  \bibinfo {pages} {97} (\bibinfo {year} {1970})}\BibitemShut {NoStop}%
\bibitem [{\citenamefont {Hu}\ and\ \citenamefont
  {Sugiyama}(1996)}]{Hu:1995en}%
  \BibitemOpen
  \bibfield  {author} {\bibinfo {author} {\bibfnamefont {W.}~\bibnamefont
  {Hu}}\ and\ \bibinfo {author} {\bibfnamefont {N.}~\bibnamefont {Sugiyama}},\
  }\href {\doibase 10.1086/177989} {\bibfield  {journal} {\bibinfo  {journal}
  {Astrophys. J.}\ }\textbf {\bibinfo {volume} {471}},\ \bibinfo {pages} {542}
  (\bibinfo {year} {1996})},\ \Eprint {http://arxiv.org/abs/astro-ph/9510117}
  {arXiv:astro-ph/9510117 [astro-ph]} \BibitemShut {NoStop}%
\bibitem [{\citenamefont {Lapi}\ and\ \citenamefont
  {Danese}(2015)}]{Lapi:2015zea}%
  \BibitemOpen
  \bibfield  {author} {\bibinfo {author} {\bibfnamefont {A.}~\bibnamefont
  {Lapi}}\ and\ \bibinfo {author} {\bibfnamefont {L.}~\bibnamefont {Danese}},\
  }\href@noop {} {\  (\bibinfo {year} {2015})},\ \Eprint
  {http://arxiv.org/abs/1508.02147} {arXiv:1508.02147 [astro-ph.CO]}
  \BibitemShut {NoStop}%
\bibitem [{\citenamefont {Amendola}(2000)}]{Amendola:1999er}%
  \BibitemOpen
  \bibfield  {author} {\bibinfo {author} {\bibfnamefont {L.}~\bibnamefont
  {Amendola}},\ }\href {\doibase 10.1103/PhysRevD.62.043511} {\bibfield
  {journal} {\bibinfo  {journal} {Phys. Rev.}\ }\textbf {\bibinfo {volume}
  {D62}},\ \bibinfo {pages} {043511} (\bibinfo {year} {2000})},\ \Eprint
  {http://arxiv.org/abs/astro-ph/9908023} {arXiv:astro-ph/9908023 [astro-ph]}
  \BibitemShut {NoStop}%
\bibitem [{\citenamefont {Zimdahl}\ and\ \citenamefont
  {Pavon}(2001)}]{Zimdahl:2001ar}%
  \BibitemOpen
  \bibfield  {author} {\bibinfo {author} {\bibfnamefont {W.}~\bibnamefont
  {Zimdahl}}\ and\ \bibinfo {author} {\bibfnamefont {D.}~\bibnamefont
  {Pavon}},\ }\href {\doibase 10.1016/S0370-2693(01)01174-1} {\bibfield
  {journal} {\bibinfo  {journal} {Phys. Lett.}\ }\textbf {\bibinfo {volume}
  {B521}},\ \bibinfo {pages} {133} (\bibinfo {year} {2001})},\ \Eprint
  {http://arxiv.org/abs/astro-ph/0105479} {arXiv:astro-ph/0105479} \BibitemShut
  {NoStop}%
\bibitem [{\citenamefont {Chimento}(2010)}]{Chimento:2009hj}%
  \BibitemOpen
  \bibfield  {author} {\bibinfo {author} {\bibfnamefont {L.~P.}\ \bibnamefont
  {Chimento}},\ }\href {\doibase 10.1103/PhysRevD.81.043525} {\bibfield
  {journal} {\bibinfo  {journal} {Phys. Rev.}\ }\textbf {\bibinfo {volume}
  {D81}},\ \bibinfo {pages} {043525} (\bibinfo {year} {2010})},\ \Eprint
  {http://arxiv.org/abs/0911.5687} {arXiv:0911.5687 [astro-ph.CO]} \BibitemShut
  {NoStop}%
\bibitem [{\citenamefont {Szydlowski}\ \emph {et~al.}(2006)\citenamefont
  {Szydlowski}, \citenamefont {Stachowiak},\ and\ \citenamefont
  {Wojtak}}]{Szydlowski:2005kv}%
  \BibitemOpen
  \bibfield  {author} {\bibinfo {author} {\bibfnamefont {M.}~\bibnamefont
  {Szydlowski}}, \bibinfo {author} {\bibfnamefont {T.}~\bibnamefont
  {Stachowiak}}, \ and\ \bibinfo {author} {\bibfnamefont {R.}~\bibnamefont
  {Wojtak}},\ }\href {\doibase 10.1103/PhysRevD.73.063516} {\bibfield
  {journal} {\bibinfo  {journal} {Phys. Rev.}\ }\textbf {\bibinfo {volume}
  {D73}},\ \bibinfo {pages} {063516} (\bibinfo {year} {2006})},\ \Eprint
  {http://arxiv.org/abs/astro-ph/0511650} {arXiv:astro-ph/0511650 [astro-ph]}
  \BibitemShut {NoStop}%
\bibitem [{\citenamefont {Zilioti}\ \emph {et~al.}(2015)\citenamefont
  {Zilioti}, \citenamefont {Santos},\ and\ \citenamefont
  {Lima}}]{Zilioti:2015rza}%
  \BibitemOpen
  \bibfield  {author} {\bibinfo {author} {\bibfnamefont {G.~J.~M.}\
  \bibnamefont {Zilioti}}, \bibinfo {author} {\bibfnamefont {R.~C.}\
  \bibnamefont {Santos}}, \ and\ \bibinfo {author} {\bibfnamefont {J.~A.~S.}\
  \bibnamefont {Lima}},\ }\href@noop {} {\  (\bibinfo {year} {2015})},\ \Eprint
  {http://arxiv.org/abs/1508.06344} {arXiv:1508.06344 [gr-qc]} \BibitemShut
  {NoStop}%
\end{thebibliography}
\end{document}